\documentclass{aa}

\usepackage{natbib}
\usepackage{graphicx}
\usepackage{txfonts}
\usepackage{hyperref}
\usepackage{pdflscape}
\usepackage{longtable}
\usepackage{color}

\newcommand{\kms}{km~s$^{-1}$}

\newcommand{\kkms}{K~km~s$^{-1}$}
\newcommand{\kkmsa}{K~km~s$^{-1}$as$^2$}
\newcommand{\gcms}{g~cm~s$^{-1}$}
\newcommand{\vlsr}{V$_\mathrm{LSR}$}
\newcommand{\vexp}{v$_\mathrm{exp}$}

\newcommand{\msun}{\mbox{M$_{\odot}$}}

\newcommand{\doce}{\mbox{$^{12}$CO\ }}
\newcommand{\trece}{\mbox{$^{13}$CO\ }}

\newcommand{\jtd}{\mbox{$J$=3$-$2}}
\newcommand{\jdu}{\mbox{$J$=2$-$1}}
\newcommand{\juc}{\mbox{$J$=1$-$0}}
\newcommand{\td}{\mbox{3$-$2}}
\newcommand{\du}{\mbox{2$-$1}}

\newcommand{\mm}{\mbox{$\mu$m}}
\def\arcdeg{\hbox{$^\circ$}}

\newcommand{\tmb}{\mbox{T$_\mathrm{mb}$}}

\begin{document}

   \title{The ionised and molecular mass of post-common-envelope planetary nebulae}

   \subtitle{The missing mass problem}

   \author{M. Santander-Garc\'ia
          \inst{1}
          \and
          D. Jones\inst{2,3}
          \and
          J. Alcolea\inst{1}
          \and
          V. Bujarrabal\inst{4}
          \and
          R. Wesson\inst{5}
          }

   \institute{Observatorio Astron\'omico Nacional (OAN-IGN), Alfonso XII, 3, 28014, Madrid, Spain,
              \email{m.santander@oan.es}
         \and
             Instituto de Astrof\'isica de Canarias, 38205, La Laguna, Spain
         \and
             Departamento de Astrof\'isica, Universidad de La Laguna, 38206, La Laguna, Spain
         \and
             Observatorio Astron\'omico Nacional (OAN-IGN), Apartado 112, 28803, Alcal\'a de Henares, Spain
        \and
            Department of Physics and Astronomy, University College London, Gower St, London WC1E 6BT, UK
             }


\authorrunning{Santander-Garc\'ia et al.}
\titlerunning{The ionised and molecular mass of post-CE PNe}


  \abstract
  {Most planetary nebulae (PNe) show beautiful, axisymmetric morphologies despite their progenitor stars being essentially spherical. Close binarity  is widely invoked to help eject an axisymmetric nebula, after a brief phase of engulfment of the secondary within the envelope of the Asymptotic Giant Branch (AGB) star, known as the common envelope (CE). The evolution of the AGB would thus be interrupted abruptly, its still quite massive envelope being rapidly ejected to form the PN, which a priori would be more massive than a PN coming from the same star, were it single.}
   {We aim at testing this hypothesis by investigating the ionised and molecular masses of a sample consisting of 21 post-CE PNe, roughly one fifth of the known total population of these objects, and comparing them to a large sample of `regular' (i.e. not known to arise from close-binary systems) PNe.}
   {We have gathered data on the ionised and molecular content of our sample from the literature, and carried out molecular observations of several previously unobserved objects. We derive the ionised and molecular masses of the sample by means of a systematic approach, using tabulated, dereddened $H\beta$ fluxes for finding the ionised mass, and \doce\ \jdu\ and \jtd\ observations for estimating the molecular mass.}
   {There is a general lack of molecular content in post-CE PNe. Our observations only reveal molecule-rich gas around NGC~6778, distributed into a low-mass, expanding equatorial ring lying beyond the ionised broken ring previously observed in this nebula. The only two other objects showing molecular content (from the literature) are NGC~2346 and NGC~7293. Once we derive the ionised and molecular masses, we find that post-CE PNe arising from Single-Degenerate (SD) systems are just as massive, on average, as the `regular' PNe sample, whereas post-CE PNe systems arising from Double-Degenerate (DD) systems are considerably more massive, and show substantially larger linear momenta and kinetic energy than SD systems and `regular' PNe. Reconstruction of the CE of four objects, for which a wealth of data on the nebulae and complete orbital parameters are available, further suggests that the mass of SD nebulae actually amounts to a very small fraction of the envelope of their progenitor stars. This leads to the uncomfortable question of where the rest of the envelope is and why we cannot detect it in the stars' vicinity, thus raising serious doubts on our understanding of these intriguing objects.}
   {}


   \maketitle
%

\section{Introduction}\label{intro}

Low- and intermediate-mass (up to $\sim$8 \msun) stars end their lives by ejecting their envelope into beautiful nebulae with intricate geometries. The resulting planetary nebulae (PNe) show high degrees of symmetry, with mostly bipolar or elliptical morphologies. The mechanism behind the shaping of axisymmetric PNe has been matter of debate for the last several decades (e.g. \citealp{balick02}), although it is becoming increasingly clear that angular momentum from a binary or sub-stellar companion is a key ingredient to this intriguing puzzle (\citealp{jones17}; \citealp{decin20}).

The close binary central stars of PNe (CSPNe) are evolved binaries with orbital separations orders of magnitude smaller than the typical radius of an Asymptotic Giant Branch (AGB) star.  As such, the component stars must have previously interacted and evolved to this separation rather than having formed thus. The mechanism, first proposed by \cite{paczynski76}, by which these systems reach their current configuration is thought to proceed in the following manner: as the primary star evolves along the giant branch(es) and expands, copiously losing mass through a slow wind, it eventually expands to overflow its Roche-lobe. Runaway mass transfer on to the secondary then occurs through the inner Lagrangian point on dynamical time scales, engulfing the companion and leading to the formation of a a common envelope (CE). In this brief ($\sim$1 year) phase, the secondary quickly spirals-in inside the extended envelope of the primary due to drag forces, leading to either a merging of the two stars, or the abrupt end and ejection of the CE \citep{ivanova13}. In the latter case, the envelope is shaped into a bipolar planetary nebula whose equator would coincide with the system's orbital plane. This is indeed the case in every one of the eight cases analysed so far, in which the orientation of the orbital plane and nebula equator could be determined (\citealp{hillwig16,munday20}). Such a  correlation constitutes the strongest statistical proof so far of the influence of close-binarity in the shaping of PNe.

The first observational confirmation of the existence of PNe with close-binary nuclei was made by \cite{bond76} in Abell 63. A few other cases came after in the following years, although the hypothesis did not really gain ground until the arrival of modern, systematic photometric surveys such as the Optical Gravitational Lensing Experiment (OGLE; \citealp{udalski08}), when dozens of close-binary central stars of PNe were detected and a solid lower limit of $\gtrsim15$\% was established for the post-CE binary fraction  \citep{miszalski09a}. The identification of morphological traits such as rings, jets and fast low-ionisation emitting regions as characteristic trends indicative of close-binarity further enhanced the statistics (e.g. \citealp{miszalski11a}, \citeyear{miszalski11b}; \citeyear{boffin12}, \citealp{jones14}, \citeyear{jones15}; \citealp{corradi11a}; \citealp{santander15b}) up until the present number of $\sim$100 confirmed binary CSPNe\footnote{See updated list with references to discovery papers in \url{http://www.drdjones.net/bcspn/}}.

On theoretical grounds, however, deeper understanding of the physics of the CE remains very elusive \citep{ivanova13,jones20}. Most hydrodynamic models are unable to gravitationally unbind the whole envelope, effectively ejecting no more than a few tenths of the whole envelope (\citealp{ohlmann16}; \citealp{ricker12}; \citealp{garcia18}), either because they lack key physical ingredients, or because of fundamental hardware limitations (see \citealp{chamandy20}). Exceptions require resorting to additional energy reservoirs, such as the recombination energy from the ionised region, which is debated (\citealp{webbink08}; \citealp{ricker12}; \citealp{nandez15}; \citealp{ivanova18}; \citealp{sand20}). In summary, although simulations collectively show the major role of the CE on the shaping of PNe, we are far from fully understanding the physics behind the death of a significant fraction of stars in the Universe.

Careful estimation of the mass of these envelopes could help provide insight in to  CE ejection through constraints that can then be fed back in to  modelling efforts. We can, in principle, derive this parameter by determining the total masses of the resultant PNe, under the assumption of sudden ejection of the CE into forming the PNe, i.e. the mass of the CE mass equals the mass of the PN ---excluding any halo, which would have been deployed into the interstellar medium (ISM) long before the CE stage. In this respect, it will also useful to put these mass figures in the context of the general population of PNe, encompassing nebulae arising not only from close binaries, but also from single stars and longer period, binary stars that did not experience a CE.

It can be argued that CE evolution implies significant differences in the mass-loss history of the central star with respect to single star evolution. Let us consider a single AGB star first. Most of the mass it loses along its evolution via slow winds gets too diluted in the ISM to be detected later during the PN stage \citep{mccullough01,villaver02}. It is instead the mass lost during the superwind phase (lasting $\sim$500-3000 years), amounting to $\sim$0.1-0.6 \msun\ for a star with an initial mass of 1.5 \msun\ (see review in \citealp{hofner18}), which will conform the PN.

Should the same AGB star be part of a close-enough binary system, its evolution will be abruptly interrupted as soon as it expands to fill its Roche-lobe, engulf its companion, and undergo the CE stage. This can be expected to occur in the final few ($\sim$1-20) million years of the AGB stage (e.g. Figure~3 in \citealp{jones20}). Thus, on the ejection of the CE, the resultant PN should, in principle, comprise all the mass the AGB star did not deploy into the ISM during these last million years (c.f.\ only the last few thousand years for the single star scenario outlined above). Due to the premature, sudden ejection of the CE, no extended, massive haloes are expected in the near vicinity of this expanding PN.

Thus, one would expect PNe arising from a CE to be more massive, on average, than their single-star and long-period binary counterparts\footnote{By this same argument, PNe arising from CE ejection during the RGB should tend to be even more massive.}.

Nevertheless, the only mass determinations of post-CE PNe available in the literature would hint toward the opposite idea. \cite{frew07} calculated the ionised masses of a sample of post-CE PNe and found them to be actually lower, on average, than those of the general PNe population \citep[a finding later reinforced by][]{corradi15}. However, \citeauthor{frew07} and \citeauthor{corradi15} only included the ionised mass in their calculations, not accounting for the potential presence of material not yet ionised or photo-dissociated by the UV radiation from the white dwarf (WD). Recent work on the dust emission around post-Red-Giant-Branch (post-RGB) stars in the LMC, thought to have undergone a CE which cut short their evolution, has indicated that the dust mass in these objects is similarly very low (10$^{-7}$--10$^{-4}$ M$_\odot$) indicating that the ``missing mass'' is not hiding in a dusty disk or shell \citep{sarkar21}. However, little is known about the molecular or neutral gas content of post-CE PNe.

With the goal of reaching a better understanding of this issue, in this work we derive the ionised and molecular mass of a sample of 21 post-CE PNe, representing roughly 20\% of the total known objects of this kind. The paper is organised as follows: in Section~\ref{sample-observations} we present the sample and describe the molecular line observations and data reduction; Section~\ref{ngc6778} deals with the mm-wavelength emission detected in our observations, and its modelling; we calculate the ionised and molecular mass of the whole sample, and compare it to regular PNe (that is, PNe not confirmed to host a close-binary) in Section~\ref{mass-postce}; finally, we discuss the results in section~\ref{discussion} and summarise our conclusions in Section~\ref{conclusions}.

\section{Sample and Observations}\label{sample-observations}

\begin{figure}
\begin{center}
 \includegraphics[width=\columnwidth]{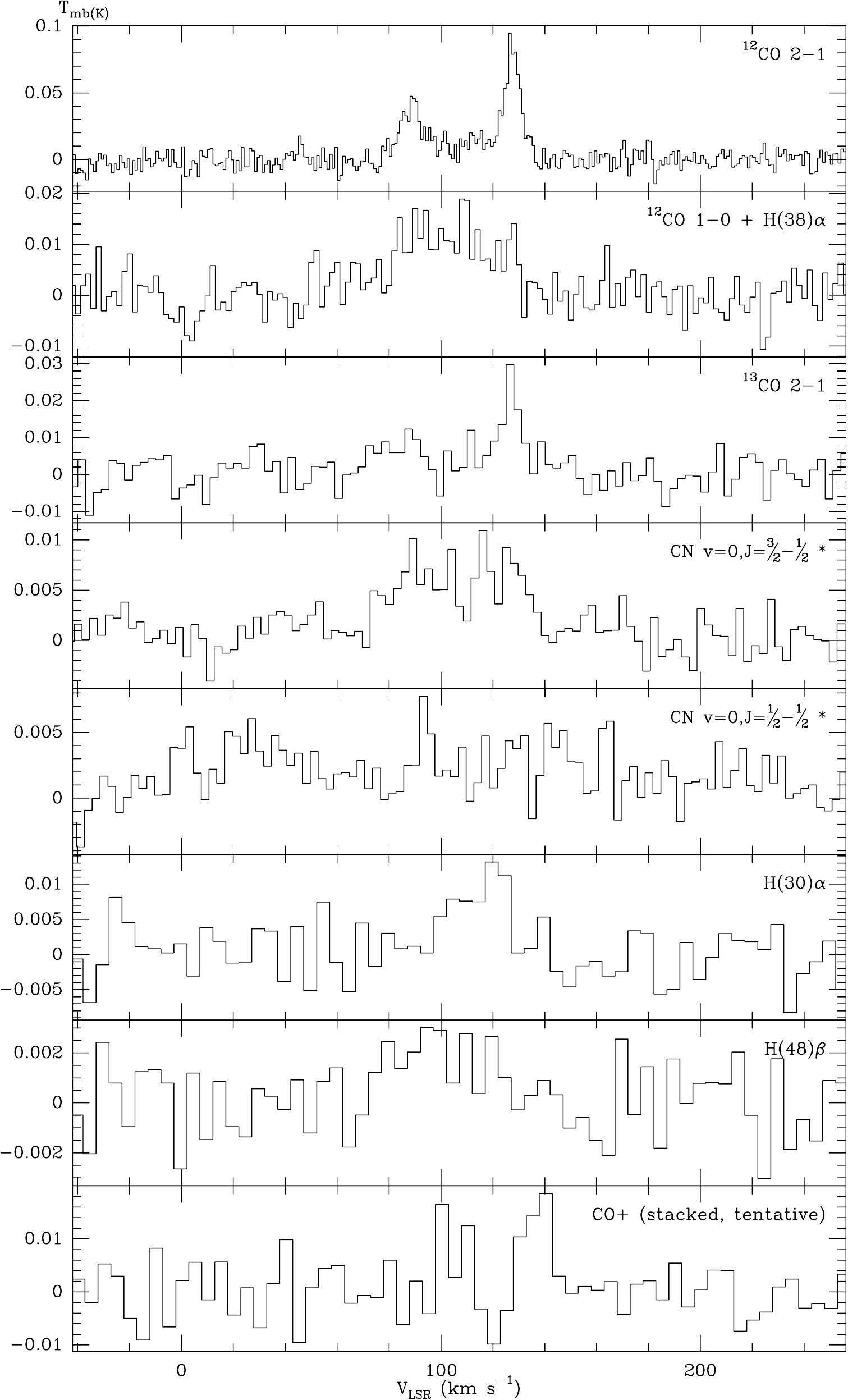}
 \caption{Detected (and tentatively detected) mm-wavelength emission at the position of the central star of NGC~6778. The systemic \vlsr\ is 107.1 \kms. ($^*$) indicates that the velocity of the CN line complexes shown, with J=3/2--1/2 and J=1/2--1/2, are referenced to frequencies of 113.490985 GHz and 113.157347 GHz, respectively. The tentative CO+ line shown is a stack from the N=2--1 group with J=3/2--3/2, 3/2--1/2, and 5/2--3/2.}
   \label{fig:ngc6778lines}
\end{center}
\end{figure}

The sample of post-CE PNe analysed in this work consists of two sub-samples. The first one consists of nine northern post-CE PNe previously unobserved in \doce and \trece\ \juc\ and \jdu, observations of which were secured with the IRAM 30m radiotelescope (see the top part of Tables~\ref{tab:undetectedsample} and \ref{tab:detectedsample} for details). These nebulae are relatively compact so as to fit inside the telescope beam in one or a few pointings, in order to account for their whole CO content as accurately as possible. This sub-sample was selected to cover a broad range of kinematical ages, orbital periods and morphologies. All of them show some emission excess in the far infrared (IR), with bumps peaking at 25-60 $\mu$m, which in evolved stars (still not undergoing ionisation) correlates with CO emission (see e.~g.\ \citealp{bujarrabal92}). Observations were carried out in two runs, in December 2017 and May 2018. The telescope half power beam width (HPBW) was 10.7 and 21.3 arcsec at 230 GHz and 110 GHz, respectively, according to the latest telescope parameters provided by IRAM. The FTS200 backends were used, and the spectral resolution degraded to 1 \kms\ in order to better detect the molecular profiles of these PNe, expected to be in the range 20-80 \kms\ wide. Data were reduced using standard baseline-subtraction and averaging procedures in the Continuum and Line Analysis Single-dish Software (CLASS) software, part of the GILDAS suite\footnote{\url{http://www.iram.fr/IRAMFR/GILDAS}}, and flux-calibrated in the main-beam (\tmb) scale. Only one PN in this subsample, NGC~6778, was detected in these observations (\doce\ \juc\  and \jdu, \trece\ \jdu). See Figure~\ref{fig:ngc6778lines} for the detected mm-wavelength emission, and Section~\ref{ngc6778} below for an analysis of the molecular emission in this nebula.

A second sub-sample was constructed from every molecular observation of PNe now confirmed to host a post-CE binary found throughout the literature (\citealp{huggins89}, \citealp{huggins96}, \citealp{huggins05}, \citealp{guzman18} and references therein). These comprise another 12 post-CE PNe, observed with either NRAO~12m, APEX~12m, SEST~15m or IRAM~30m radiotelescopes, in different configurations including \doce \juc, and/or \jdu, or \jtd. The only exclusion in this work is that of HaTr~4, observed (and undetected) by \citealp{guzman18}, because no integrated, dereddened H$\beta$ (or H$\alpha$) flux  could be found in the literature. The bottom part of Tables~\ref{tab:undetectedsample} and \ref{tab:detectedsample} summarise these observations, providing the transitions used, the velocity resolution, the rms achieved around the undetected line or the integrated flux, depending on the case. These figures were extracted directly from \citealp{huggins89}, \citealp{huggins96}, while data from \citealp{guzman18} were reanalysed from APEX archival data, since these authors did not provide the velocity resolution corresponding to their quoted sensitivities. The resulting data of this subsample follow a similar pattern as our own observations: only two objects, NGC~2346 and NGC~7293, show molecular emission down to the different sensitivities achieved.

\begin{table*}[h]
\centering
 \caption[]{\label{mm-nondetect}Post-CE PNe undetected at mm-wavelengths.}
\begin{tabular}{llllcccl}
 \hline \hline
  Nebula &
  PN~G &
  Telescope &
  Transition &
  HPBW &
  $\Delta$v resol. &
  rms &
  Notes  \\
   &  &  &  & (arcsec) & (\kms) & (mK) & \\
 \hline
\noalign{\smallskip}
\multicolumn{8}{l}{{\bf Observed sub-sample}} \\
\noalign{\smallskip}
\hline
Abell~41  & G009.6+10.5 & IRAM~30m & \doce \juc & 21.3 & 1.0 & 12  &  \\
          &            &          & \doce \jdu & 10.7 & 1.0 & 8.2 & \\
Hen~2-428 & G049.4+02.4 & IRAM~30m & \doce \juc & 21.3 & 1.0 & 9.7 &  \\
          &            &          & \doce \jdu & 10.7 & 1.0 & 6.8 & \\
Abell~63  & G053.8-03.0 & IRAM~30m & \doce \juc & 21.3 & 1.0 & 16  & \\
          &            &          & \doce \jdu & 10.7 & 1.0 & 9   & \\
Necklace  & G054.2-03.4 & IRAM~30m & \doce \juc & 21.3 & 1.0 & 10  & 1 \\
          &            &          & \doce \jdu & 10.7 & 1.0 & 8.5 & 1 \\
V458~Vul  & G058.6-03.6 & IRAM~30m & \doce \juc & 21.3 & 1.0 & 9.9 &  \\
          &            &          & \doce \jdu & 10.7 & 1.0 & 8.1 & \\
ETHOS~1   & G068.1+11.0 & IRAM~30m & \doce \juc & 21.3 & 1.0 & 11  & 2 \\
          &            &          & \doce \jdu & 10.7 & 1.0 & 7.2 & 2 \\
Ou~5      & G086.9-03.4 & IRAM~30m & \doce \juc & 21.3 & 1.0 & 12  & \\
          &            &          & \doce \jdu & 10.7 & 1.0 & 10  & \\
PM~1-23   & G221.8-04.2 & IRAM~30m & \doce \juc & 21.3 & 1.0 & 48  &  \\
          &            &          & \doce \jdu & 10.7 & 1.0 & 100 & \\
\hline
\noalign{\smallskip}
\multicolumn{8}{l}{{\bf Sub-sample from the literature}} \\
\noalign{\smallskip}
\hline
NGC~246   & G118.8-74.7 & NRAO~12m & \doce \jdu & 31 & 1.3 & 62    & 3 \\
NGC~2392  & G197.8+17.3 & NRAO~12m & \doce \jdu & 31 & 1.3 & 57    & 3 \\
Abell~30  & G208.5+33.2 & NRAO~12m & \doce \jdu & 31 & 1.3 & 47    & 3 \\
Fg~1      & G290.5+07.9 & SEST~15m & \doce \jdu & 24 & 0.9 & 48    & 4 \\
NGC~5189  & G307.2-03.4 & SEST~15m & \doce \jdu & 24 & 0.9 & 82    & 4 \\
MyCn~18   & G307.5-04.9 & APEX~12m & \doce \jtd & 18 & 0.066 & 149 & 5 \\
NGC~6326  & G338.1-08.3 & APEX~12m & \doce \jtd & 18 & 0.066 & 68  & 5,6 \\
Hen~2-155 & G338.8+05.6 & SEST~15m & \doce \jdu & 24 & 0.9 & 42    & 4 \\
Sp~3      & G342.5-14.3 & SEST~15m & \doce \jdu & 24 & 0.9 & 52    & 4 \\
Lo~16     & G349.3-04.2 & APEX~12m & \doce \jtd & 18 & 0.066 & 52  & 5 \\
\hline
\end{tabular}
\tablefoot{(1)~Position to the NW (by the largest knot visible in H$\alpha$, offset $\sim$6 arcsec from the central star) observed at similar noise; (2) North cap (H$\alpha$ peak, offset 29 arcsec from the central star) observed at similar noise; (3) Data from \citet{huggins89}; (4) Data from \citet{huggins96};
(5) Data from \citet{guzman18}, re-analysed from archive; (6) (Tentative) detection in original work not observed in our re-analysis.}
\label{tab:undetectedsample}
\end{table*}

\begin{table*}[h]
\centering
 \caption[]{\label{mm-detect}Post-CE PNe detected at mm-wavelengths.}
\begin{tabular}{llllcccl}
 \hline \hline
  Nebula &
  PN~G &
  Telescope &
  Transition &
  HPBW &
  $\Delta$v resol. &
  Intensity &
  Notes  \\
   &  &  &  & (arcsec) & (\kms) & (K~\kms) & \\
 \hline
\noalign{\smallskip}
\multicolumn{8}{l}{{\bf Observed sub-sample}} \\
\noalign{\smallskip}
\hline
NGC~6778  & 034.5-06.7 & IRAM~30m & \doce \juc + H(38)$\alpha$ & 21.3 & 1.0 & 0.58 &  1 \\
          &            &          & \doce \jdu & 10.7 & 1.0 & 1.4  &  1 \\
          &            &          & \trece \jdu & 11.2 & 1.0 & 0.5 & 1 \\
          &            &          & CN v=0, J=3/2--1/2 & 21.7 & 1.0 & 0.18  &  1 \\
          &            &          & CN v=0, J=1/2--1/2 & 21.7 & 1.0 & 0.14  &  1 \\
          &            &          & H(30)$\alpha$ & 10.6 & 1.0 & 0.3  &  1 \\
          &            &          & H(48)$\beta$ & 22.0 & 1.0 & 0.07  &  1 \\
          &            &          & CO+ N=2--1 & 10.4 & 1.0 & 0.29  &  Tentative, 3-line stack\\

\hline
\noalign{\smallskip}
\multicolumn{8}{l}{{\bf Sub-sample from the literature}} \\
\noalign{\smallskip}
\hline
NGC~7293  & 036.1-57.1 & NRAO~12m &  \doce \jdu & 31 & 0.65 & 13.2  & 2\\
NGC~2346  & 215.6+03.6 & IRAM~30m &  \doce \jdu & 12 & 1.3 & 21  & 3\\
\hline
\end{tabular}
\tablefoot{(1)~Additional positions offset 10 arcsec to the east and west along the nebula equator (Position Angle 114\arcdeg), and 12.5 and 14 arcsec to the north and south, respectively, along the major axis (P.A.~24\arcdeg), observed at similar noise. Flux computed from assumed size as described in Section~\ref{ngc6778}; (2)~Data from \citet{huggins86}; (3)~Data from \citet{huggins96}.}
\label{tab:detectedsample}
\end{table*}

\section{mm-wavelength emission from NGC 6778}\label{ngc6778}

The only detection in our observations with the IRAM~30m telescope was that of NGC~6778. The flux-calibrated (\tmb\ scale) profiles of detected transitions at the position of the central star system are displayed in Fig.~\ref{fig:ngc6778lines}. Observations are compatible with a systemic \vlsr\ of 107.1 \kms.

In addition to \doce\ and \trece, the CN v=0 J=3/2--1/2 and J=1/2--1/2 system, as well as the Hydrogen recombination lines H(30)$\alpha$ and H(48)$\beta$ were detected. These lines are displayed in Fig.~\ref{fig:ngc6778lines}, while their integrated intensities are shown in Table~\ref{tab:detectedsample}. CO+ (N=2--1) is also tentatively detected, with its most prominent component peaking at a velocity of $\sim$+13 \kms\ redward of the systemic velocity, which is considerably faster than the peaks seen in \doce, but still below the $\sim$26 \kms\ velocity displayed by the material in the optical range (\citealp{guerrero12}), which would hint towards the existence of a region between the molecule-rich and the ion-rich regions, where CO could be ionised by UV photons from the central star, should the tentative detection of CO+ be confirmed in this source.

\subsection{\doce\ and \trece\ in NGC~6778}\label{ngc6778-co}

\begin{figure}
\begin{center}
 \includegraphics[width=\columnwidth]{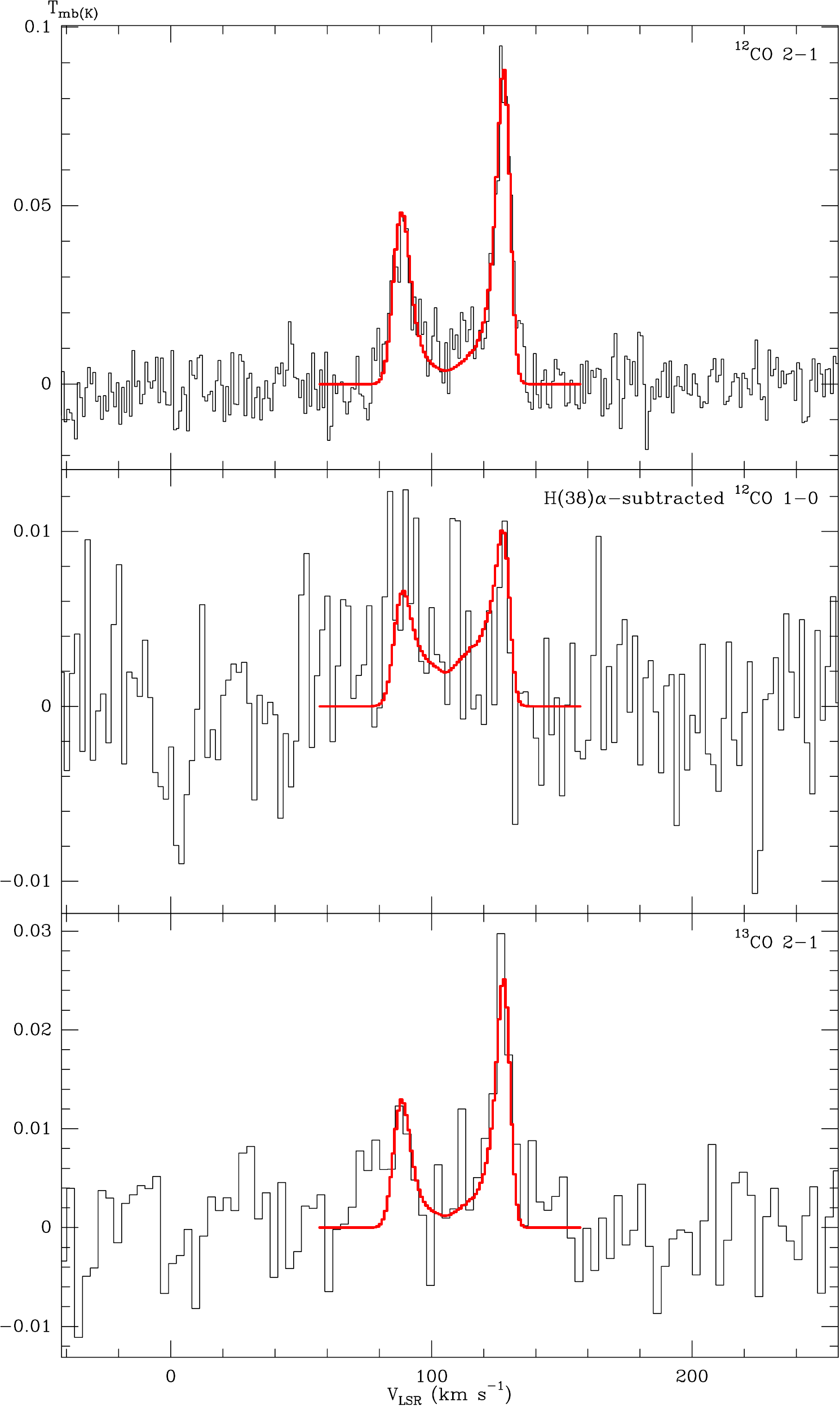}
 \caption{Detected \doce\ and \trece\ emission profiles at the position of the central star system of NGC~6778 (black), and corresponding {\tt SHAPE+SHAPEMOL} model synthetic emission profiles (red). A gaussian profile simulating H(38)$\alpha$ has been subtracted from the \doce\ \juc\ emission in order to account from contamination from this recombination line (see text).}
   \label{fig:model-central}
\end{center}
\end{figure}

The \doce \juc\ profile is contaminated by close H(38)$\alpha$ emission at 115274.41~MHz. In order to account for this contamination, we computed the relative fluxes of different Hydrogen recombination lines in IRAM~30m survey spectra of one of the best studied PNe, NGC~7027. Assuming similar physical conditions for the ionised component of NGC~6778, we concluded the intensity of H(38)$\alpha$ to be 0.65 times that of the detected H(30)$\alpha$ line (see fig.~\ref{fig:ngc6778lines}) at 231900.928 MHz. Thus we used a scaled-down H(30)$\alpha$ profile as a template for subtracting the H(38)$\alpha$ from the \doce \juc\ spectral profile, resulting in the middle panel of Fig.~\ref{fig:model-central}.

\begin{figure*}
\begin{center}
 \includegraphics[width=15cm]{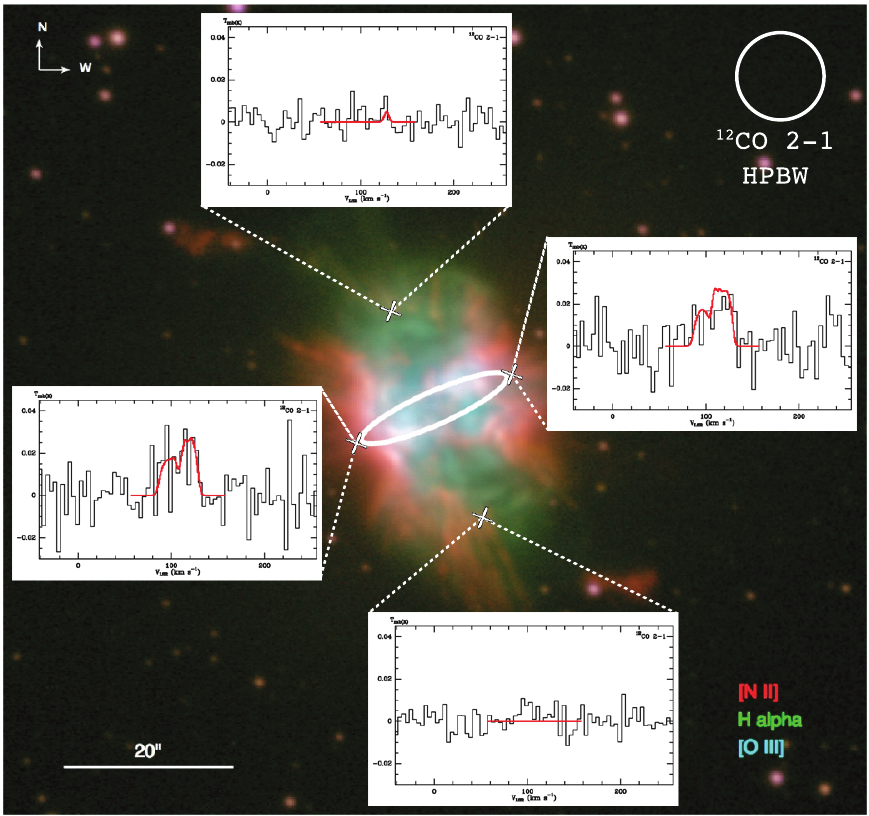}
 \caption{Equatorial ring model (white) of the \doce\ and \trece\ emission of NGC~6778 overlaid on an image of the nebula taken with the NOT telescope (adapted from \citealp{guerrero12}). White crosses mark the observed offset positions, while associated insets show their corresponding \doce\ \jdu\ emission profiles (black) and {\tt SHAPE+SHAPEMOL} model synthetic profiles (red). The Half Power Beam Width (HPBW) of the IRAM 30m telescope at the \doce\ \jdu\ transition is indicated by the white circle.}
   \label{fig:model-offset}
\end{center}
\end{figure*}

The detected \doce and \trece spectral profiles are double-peaked, with peak velocities similar to (although slightly lower than) those found by \cite{guerrero12} in the [N {\sc ii}]-emitting equatorial, distorted ring. The CO-rich domain of this nebula seems to be constrained to the central region, judging from the substantial emission decrease when offsetting the telescope by 10 arcsec along the equatorial direction, and the sharp drop at positions 12.5 and 14 arcsec away along the nebular axis (see Fig.~\ref{fig:model-offset}). With respect to peak intensities, the blue peak is fainter both in \doce and \trece. In fact, the peak-to-peak ratio is larger in \trece than it is in \doce, thus ruling out self absorption, and pointing towards a clumpy, inhomogeneous matter distribution.

We have therefore interpreted this structure as a thin equatorial ring with an approximate projected size of 14$\times$7.5 arcsec$^2$ (and thus an inclination of $\sim$32$^\circ$ to the line of sight), which we use for computing the molecular mass later, in Section~\ref{mass-postce}. We have built a spatio-kinematical model including radiative transfer in CO lines under the Large Velocity Gradient (LVG) assumption, by making use of the {\tt SHAPE+shapemol} code (\citealp{steffen11}; \citealp{santander15a}). Given the limited amount of geometric information available and the blueward and redward peak differences,  for the sake of simplicity the model is split into two semi-torus, receding from and approaching to the observer, respectively.

The achieved best-fit is shown in red in Figs.~\ref{fig:model-central} and \ref{fig:model-offset}, and the corresponding parameters, along with uncertainties (as estimated by varying each individual parameters until a fair fit is no longer achieved), are provided in Table~\ref{tab:ngc6778model}. The characteristic microturbulence velocities of the receding and approaching structures were found to be of 3 and 4 \kms, respectively.

We found the \doce to \trece abundance ratio to be as low as 4, although typical calibration errors ranging from 10 to 20\% would allow for somewhat larger abundance ratios. In any case, it is worth noting that such a low isotopic ratio could indicate an O-rich nature of this source (e.g. \citealp{milam09}).

The density, volume and \doce abundance found for this structure allows for a molecular mass estimate independent from the method followed in Section~\ref{mass-postce}. With the assumption that the bulk of the mass consists of Hydrogen molecules, and an additional correction factor of 1.2 to account for helium abundance (assumed to be H/He=0.1), the resulting molecular mass of NGC~6778 is 1.1$\times$10$^{-2}$ \msun. This figure is compatible within errors with the molecular mass found in Section~\ref{mass-postce} for this object, 0.024$\pm$0.02~\msun. Note that, given the apparent clumpy nature of this equatorial ring, the actual mass could be somewhat larger due to opacity being larger than modelled in this Section.

\begin{table*}\renewcommand{\arraystretch}{1.3}
  \begin{center}
  \caption{Best-fit model parameters for the molecular component of NGC~6778.}
  \label{tab:ngc6778model}
  \begin{tabular}{|l|c|c|c|c|c|c|c|}\hline
Component & $r_\mathrm{inner}$ &  $r_\mathrm{outer}$  &  $V_\mathrm{exp}$ & $X {\mathrm{(}^{12}\mathrm{CO)}}$ & $X {\mathrm{(}^{13}\mathrm{CO)}}$ & $n$ & $T$  \\
 & (10$^{17}$~cm) & 10$^{17}$~cm)   &   (km~s$^{-1}$) &  &   & (cm$^{-3}$) & (K)  \\
\hline
Receding semi-torus & 3.8$^{+0.01}_{-0.01}$ & 4.03$^{+0.01}_{-0.01}$  & 22.0$^{+1.0}_{-1.0}$ & 8$^{+1}_{-1}\times$10$^{-5}$ & 2$^{+0.1}_{-0.1}\times$10$^{-5}$ & 7$^{+1}_{-1}\times$10$^{3}$ & 50$^{+20}_{-10}$ \\
Approaching semi-torus & 3.8$^{+0.02}_{-0.02}$ & 4.03$^{+0.01}_{-0.01}$ &  20.0$^{+2.0}_{-1.0}$ & 8$^{+1}_{-1}\times$10$^{-5}$ & 2$^{+0.1}_{-0.3}\times$10$^{-5}$ & 3.75$^{+0.55}_{-0.65}\times$10$^{3}$ & 50$^{+20}_{-20}$ \\
\hline
  \end{tabular}
 \end{center}
\vspace{1mm}

\end{table*}

\subsection{CN in NGC 6778}\label{ngc6778-cn}

\begin{figure}
\begin{center}
 \includegraphics[width=\columnwidth]{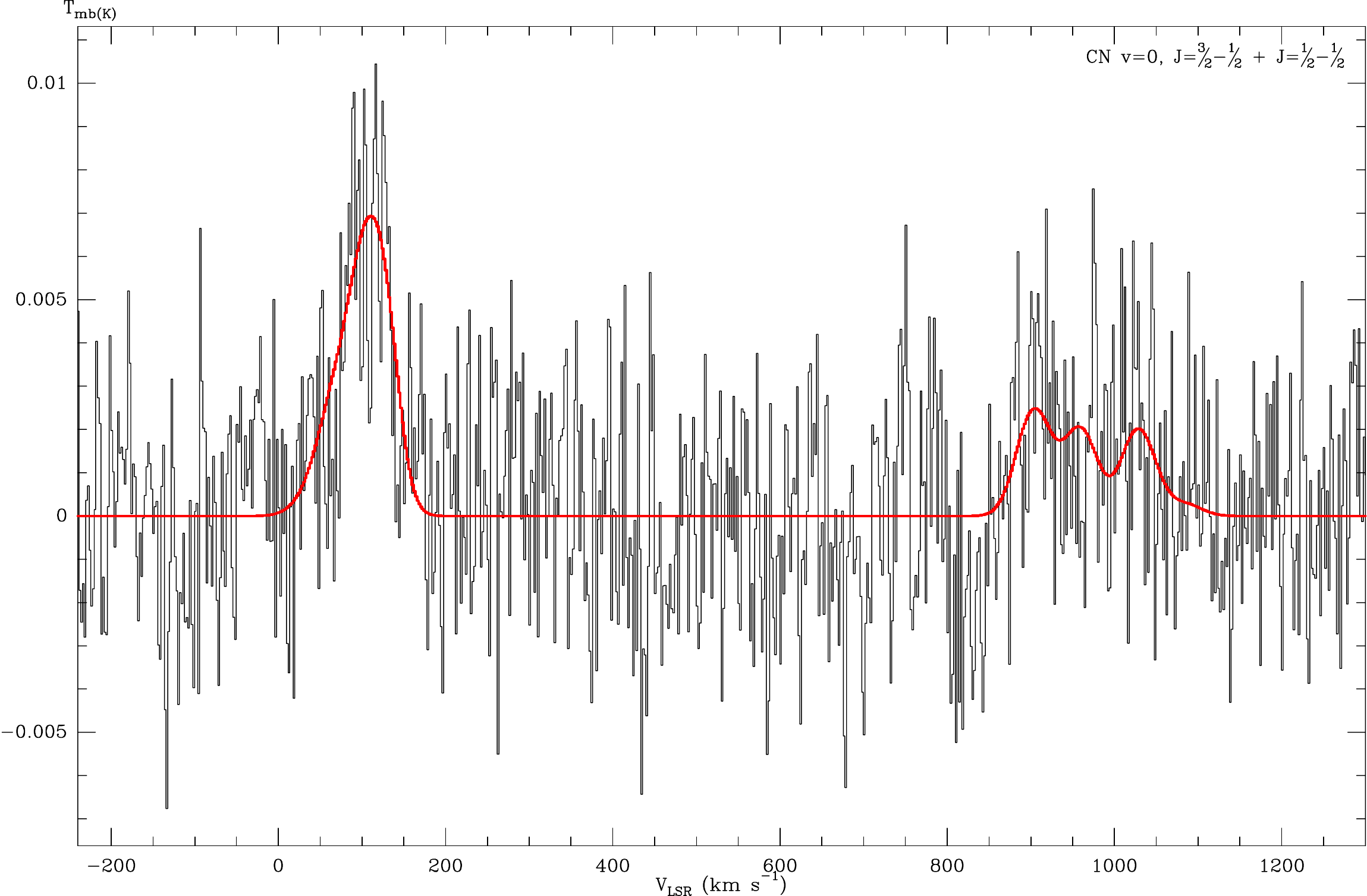}
 \caption{CN v=0 J=3/2--1/2 and J=1/2--1/2 line complexes in NGC 6778. The red line indicates hyperfine structure modelling of CN emission (see Section~\ref{ngc6778-cn}).}
   \label{fig:model-cn}
\end{center}
\end{figure}

We have modelled the detected CN emission using the hyperfine-structure-dedicated {\tt CLASS} method. The result is displayed in Figure~\ref{fig:model-cn}. The resulting optical depth of the main component is 0.7$\pm$0.4. Alternatively, the proportion existing between the intensities and integrated areas of the J=3/2--1/2 and J=1/2--1/2 groups hints towards the CN lines being rather optically thin, and thus the excitation temperature is probably rather low.

We have further investigated the abundance of CN by means of simple modelling. Assuming LTE conditions and that the spatial distribution of CN is similar to that of CO, we can compare column densities resulting from simple LTE modelling of the nearby \doce\ \juc, and \trece\ \jdu\, both of which we assume to be optically thin and at a temperature of 50~K, as found in section~\ref{ngc6778-co}. We have considered two models for CN, with temperatures of 10~K and 50~K, respectively. Using the \doce\ and \trece\ abundances found in section~\ref{ngc6778-co}, we arrive at a range of CN abundances $X_\mathrm{CN}$ between 3$\times$10$^{-7}$ and 8$\times$10$^{-7}$ when comparing with \doce\ \juc, and between 3$\times$10$^{-7}$ and 1$\times$10$^{-6}$ when comparing with \trece\ \jdu. Despite the obvious lack of information on the CN emission properties, our CN relative abundance estimates  are remarkably coherent. The relative high abundance deduced for NGC~6778 is within the range of values found in PNe (e.g. \citeauthor{bachiller97a}, \citeyear{bachiller97a,bachiller97b}), somewhat higher than abundances in O-rich AGB stars but lower than in C-rich AGBs. This probably reflects a somewhat rich chemistry taking place in a photodissociation region (PDR), something characteristic of PNe rather than of circumstellar envelopes of giant stars.

\section{The mass of post-CE PNe}\label{mass-postce}

We derived the ionised and molecular masses of the sample of post-CE PNe in a systematic manner in order to find patterns and correlations that can inform models of the CE. In this section we describe the analyses performed on both ionised and molecular components, discuss their scope and limitations, present the results, and provide some context by comparing them to the masses of a large sample of PNe derived in the same fashion.

\subsection{Ionised masses}
\label{ionisedmass-postce}

The total ionised masses of the PNe in the sample were calculated using the relation:

\begin{equation}
M_\mathrm{ion} = \frac{4\ \pi\ D^2\ F(\mathrm{H}\beta)\ m_p}{h\nu_{\mathrm{H}\beta}\ n_e\ \alpha^{\mathrm{eff}}_{\mathrm{H}\beta}},
\end{equation}

where $D$ is the distance, $F(\mathrm{H}\beta)$ is the dereddened, spatially integrated H$\beta$ flux, $m_p$ is the mass of the proton, $h\nu_{\mathrm{H}\beta}$ is the energy of an H$\beta$ photon, $n_e$ is the electron density, and $\alpha^{\mathrm{eff}}_{\mathrm{H}\beta}$ is the effective recombination coefficient of H$\beta$ \citep{corradi15}.

In order for results to be as standardised as possible, we almost exclusively used $F(\mathrm{H}\beta)$ fluxes derived from the dereddened $S(\mathrm{H}\alpha)$ surface brightness tabulated by \cite{frew16}, integrated over the ellipse defined by the minor and major axes tabulated by the same authors. We also used T[O~\textsc{III}] determinations for the electron temperatures, except in those cases without available data, where we assumed $T_e$=10\,000 K. As for electron densities, we almost exclusively used determinations based on the [S~\textsc{ii}] line doublet, except in the case of NGC~246, where only an estimate based on [O~\textsc{ii}] was available.  Finally, with respect to the distances to the objects, we prioritised GAIA eDR3 determinations by \cite{gaia20} as long as they both matched identifications by \citeauthor{chornay20} (\citeyear{chornay20,chornay21}), and their associated errors were $<33\%$. In the absence of these, we used distances by \cite{frew16}, or distance determinations to particular objects available in the literature, should the former be also absent. Every parameter used in this analysis, along with its reference, is shown in table \ref{tab:parameters}.

In each case, the ionised mass and corresponding uncertainty was derived using via 100\,000 Monte Carlo samples of the distance, electron temperature, electron density and H$\beta$ flux. A normal probability distribution was used for distances quoted with symmetric uncertainties, while for distances with asymmetric uncertainties the probability distribution was assumed to be log-normal. Where no uncertainty was available, a normal distribution was employed corresponding to an uncertainty of $\pm$20\%.  For electron temperature, again a normal distribution was employed and with an uncertainty of $\pm$5\% assumed in cases where no uncertainty was available in the literature. For the electron density, a log-normal probability distribution was assumed as this found to be the best representation of the distribution based on a random sampling of a gaussian distribution for the underlying emission line ratio, [S~\textsc{ii}]$\lambda\lambda$ 6716\AA{}/6732\AA{} \citep{wesson12}.  Where no uncertainty on the density was available, an uncertainty on the emission line ratio of 0.2 was assumed and propagated through to the derived density uncertainty.  The effective recombination coefficient of H$\beta$ was calculated using the relationship of \citet{storey95}, taking into account the dependence on both electron density and temperature.

\subsection{Molecular masses}
\label{molecularmass-postce}

Only three objects in our sample show molecular emission in either our observations (NGC~6778, see section~\ref{ngc6778}) or the data available in the literature (NGC~7293, \citealp{huggins86}; NGC~2346, \citealp{huggins96}). We therefore derived the molecular mass for these three objects, as well as conservative (3-$\sigma$) upper limits to the molecular mass of the rest of the sample based on the sensitivities achieved.

The method for estimating the molecular mass of these PNe from their CO emission relies on several simple assumptions: \textit{i)} the CO level populations are in Local Thermodynamic Equilibrium (LTE), and thus it can be characterised by a single excitation temperature $T_\mathrm{ex}$; \textit{ii)} the CO abundance $X$ relative to Hydrogen is constant throughout the molecule-rich nebula; and \textit{iii)} the selected CO transition is optically thin. Conditions \textit{i)} and \textit{ii)} are very probably satisfied in molecule-rich components, because of the favorable excitation and chemical conditions of CO (see e.g. \citealp{huggins96}, \citealp{bujarrabal01}). \textit{iii)} is discussed below. These three conditions being fulfilled, the total molecular mass $M_\mathrm{mol}$ of a nebula is:

\begin{equation}
M_\mathrm{mol} = \frac{4\ \pi\ m_\mathrm{H_2}\ D^2}{A_\mathrm{ul}\ X\ h\ \nu\ g_u}\ e^\frac{h \nu}{k T_\mathrm{ex}}\ Z(T_\mathrm{ex})\ f_\mathrm{He}\ S_\nu,
\end{equation}

where $m_\mathrm{H_2}$ is the mass of the Hydrogen molecule, $h$ and $k$ are the Planck and Boltzmann constants respectively, $\nu$ is the frequency of the transition, $A_\mathrm{ul}$ its Einstein coefficient, $g_u$ the degeneracy of its upper state, $Z$ the partition function, $D$ the distance to the nebula, f$_\mathrm{He}$ the correction factor to account for helium abundance (assumed to be He/H=0.1 and thus resulting in f$_\mathrm{He}$=1.2, since we also assume the majority of particles to be of molecular hydrogen), and $S_\nu$ the flux density of the transition, which in turn is:

\begin{equation}
S_\nu = \frac{2\ k\ \nu^2\ F}{c^2},
\end{equation}

where $c$ is the speed of light in vacuum, and $F$ the total flux of the nebula in the given transition, integrated both spatially and spectrally.

We computed the molecular masses of our sample following this scheme, assuming an excitation temperature of $T_\mathrm{ex}$=50~K, and a CO abundance $X$=2$\times$10$^{-4}$ for every object. The majority of the data of CO emission of our sample (as well as in general of PNe) available in the literature consists of surveys of \doce\ \jdu, and sometimes of the weaker \doce\ \juc\ at fairly low sensitivities, such as those by \cite{huggins89}, \cite{huggins96}, and \cite{huggins05}. Data for a few objects comes instead from a survey of \doce\ \jtd\ emission (\citealp{guzman18}).

It has been noted that both the \doce\ \jdu\ and \jtd\ lines are often optically thick to some degree in PNe, thus resulting in molecular masses in those studies being underestimated. While \trece\ \juc\ and \jdu\ emission is generally optically thin in PNe, and thus would warrant accurate mass determinations, their detection is much more difficult given their relatively low intensities, and hence data from these lines are very scarce in the literature.

In order to overcome this limitation and provide systematic, statistically meaningful, yet simple estimates of molecular masses of the sample of post-CE PNe, we opted for the following approach. Our observations of NGC~6778 plus a literature search reveal 7 PNe with both \doce \juc\ and \jdu\ detected emission, and 6 more with both \doce \jdu\ and \jtd\ detected emission (\citealp{huggins96}, \citealp{huggins05}, \citealp{guzman18}). We thus computed their masses according to each of the transitions, and computed the average correction factor needed to correct the underestimated masses resulting from \jdu\ and \jtd\ transitions, in order to match masses found via the \juc\ transition. These resulted in a factor 3.65 to be applied to calculations using \doce\ \jdu\, and a factor 5.0 for those using \doce \jtd. We therefore apply these correction factors to every PN of the sample in our molecular mass estimates. Although the validity of these correction factors will vary on an object by object basis, depending on its particular physical conditions (as other assumed values, such as the excitation temperature and CO abundance, indeed do), such a systematic correction allows for statistical comparisons with the ionised mass of these objects, and among sub-classes of post-CE PNe.

In the two cases where \doce\ emission is detected and mapping measurements are available (NGC~2346 and NGC~7293), we used those as the flux value $F$. In the case of NGC~6778, we estimated the flux by assuming constant surface brightness over the whole nebula. Thus, we integrated the detected intensity over an ellipse with major axes as estimated in Section~\ref{ngc6778}, coupled with the telescope beam. For the rest the sample, we used the 3-$\sigma$ sensitivities achieved to infer an upper limit to the intensity $I$. In order to derive the corresponding flux $F$, we integrated $I$ spatially over the ellipse defined by the nebular major and minor axes tabulated by \cite{frew16} and coupled with the telescope beam, and spectrally over an assumed velocity width of 45~\kms\ wherever the telescope beam was larger than the nebular average diameter, and of 45$\times\frac{beam}{diameter}$~\kms\ (down to a minimum of 3~\kms) wherever the beam was smaller than the nebula (with its diameter defined as the mean of its axes), thus following the same strategy as \cite{huggins96}. Note that this approach is unlikely to underestimate the molecular mass of the nebulae, since the coupling of the telescope beam with an ellipse of constant surface brightness and size as large as the optical nebula systematically results in a flux  larger than the one resulting from single-dish mapping, for 13 out of 15 nebula in which both measurements are available (\citealp{huggins96}), with these excesses having a geometric mean of 2.4. Finally, for the criteria followed for selecting the distances to the objects, see the section~\ref{ionisedmass-postce}.

Calculated errors correspond to formal error propagation. Parameters including formal errors are the distance, the correction factors discussed above (for which we take their standard deviation as error), and a 20\% relative error in intensities and fluxes to account for telescope calibration uncertainties.

Every parameter used in our analysis are displayed in Table~\ref{tab:parameters}

\subsection{Results}\label{results}

\begin{table*}
\caption{Computed ionised and molecular masses of the post-CE sample. Masses as determined here scale with distance squared. \label{tab:postcemass}}             
\centering                          
\begin{tabular}{l l c c c}        
\hline\hline                 
PN G & Common name & $D$ & M$_\mathrm{ion}$ & M$_\mathrm{mol}$ \\
    &             &  (kpc)   &    (\msun)         & (\msun)  \\
\hline\hline
\multicolumn{5}{c}{\sc{Single-Degenerate post-CE PNe}}\\
\hline
G034.5-06.7 & NGC~6778 & 2.79$\pm$0.79 & 0.19$^{+0.14}_{-0.10}$ & 0.02$\pm$0.02 \\
G036.1-57.1 & NGC~7293 & 0.200$\pm$0.002 & 0.09$^{+0.13}_{-0.05}$ & 0.3$\pm$0.2 \\
G053.8-03.0 & Abell~63 & 2.703$\pm$0.219 & 0.012$^{+0.04}_{-0.009}$ & $<$0.006 \\
G054.2-03.4 & Necklace & 4.6$\pm$1.1 & 0.009$^{+0.017}_{-0.006}$ & $<$0.007 \\
G068.1+11.1 & ETHOS~1 & 4.2$\pm$0.0 & 0.008$^{+0.3}_{-0.008}$ & $<$0.007 \\
G086.9-03.4 & Ou~5 & 5.0$\pm$1.0 & 0.18$^{+0.4}_{-0.12}$ & $<$0.012 \\
G118.8-74.7 & NGC~246 & 0.556$\pm$0.025 & 0.07$^{+0.12}_{-0.05}$ & $<$0.02 \\
G208.5+33.2 & Abell~30 & 2.222$\pm$0.148 & 0.015$^{+0.02}_{-0.009}$ & $<$0.20 \\
G215.6+03.6 & NGC~2346 & 1.389$\pm$0.039 & 0.09$^{+0.09}_{-0.04}$ & 0.7$\pm$0.5 \\
G221.8-04.2 & PM~1-23 & 5.2$\pm$2.0 & 0.015$^{+1.2}_{-0.014}$ & $<$0.17 \\
G307.5-04.9 & MyCn~18 & 4.000$\pm$1.280 & 0.07$^{+0.10}_{-0.04}$ & $<$0.06 \\
G338.1-08.3 & NGC~6326 & 5.000$\pm$1.500 & 0.6$^{+0.5}_{-0.3}$ & $<$0.06 \\
G338.8+05.6 & Hen~2-155 & 4.348$\pm$1.323 & 0.3$^{+0.2}_{-0.14}$ & $<$0.10 \\
G342.5-14.3 & Sp~3 & 2.22$^{+0.61}_{-0.48}$ & 0.09$^{+0.08}_{-0.04}$ & $<$0.06 \\
G349.3-04.2 & Lo~16 & 1.818$\pm$0.132 & 0.4$^{+0.7}_{-0.3}$ & $<$0.013 \\
\hline
\multicolumn{5}{c}{\sc{Double-Degenerate post-CE PNe}}\\
\hline
G009.6+10.5 & Abell~41 & 4.89$\pm$1.4 & 0.16$^{+0.15}_{-0.09}$ & $<$0.011 \\
G049.4+02.4 & Hen~2-428 & 4.545$\pm$1.446 & 0.7$^{+0.8}_{-0.4}$ & $<$0.010 \\
G058.6-03.6 & V458~Vul & 12.5$\pm$2.0 & 0.11$^{+0.19}_{-0.07}$ & $<$0.08 \\
G197.8+17.3 & NGC~2392 & 1.818$\pm$0.165 & 0.4$^{+0.4}_{-0.19}$ & $<$0.09 \\
G290.5+07.9 & Fg~1 & 2.564$\pm$0.197 & 0.4$^{+0.2}_{-0.15}$ & $<$0.09 \\
G307.2-03.4 & NGC~5189 & 1.471$\pm$0.043 & 0.11$^{+0.03}_{-0.03}$ & $<$0.09 \\
\hline
\end{tabular}
\end{table*}

\begin{figure}
\begin{center}
 \includegraphics[width=\columnwidth]{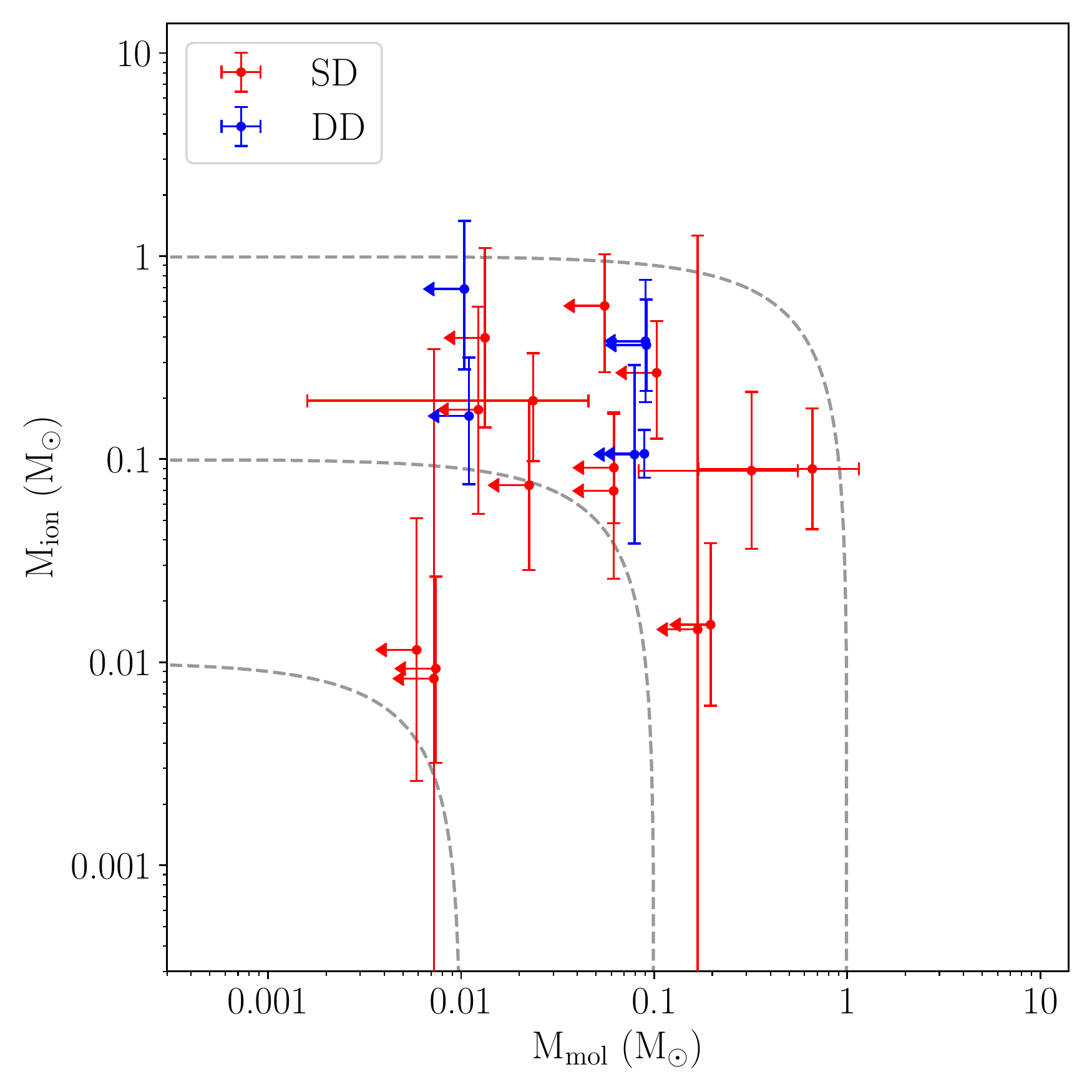}
 \caption{Ionised vs. molecular mass of our post-CE PNe sample. The further to the top and to the right a nebula is, the more massive it is. Dashed lines represent `isomasses', indicating equal ionised+molecular mass; neglecting neutral atomic mass (see Section~\ref{results}), individual nebulae run along these lines as their gas content is progressively ionised.}
   \label{fig:ionvsmolmass-postce}
\end{center}
\end{figure}

The ionised and molecular masses found in this work for the analysed sample of post-PNe are shown in Table~\ref{tab:postcemass}. An interesting trend arises when dividing the sample into two categories, namely Single-Degenerate (SD) and Double-Degenerate (DD) systems, according respectively to one or both components of the binary pair being post-AGB stars. Thus, PNe hosting DD systems seem substantially more massive than those hosting SD systems. The ionised and molecular masses of the whole sample are displayed in Fig.~\ref{fig:ionvsmolmass-postce}.

Note that the analysis presented here does not take into account any mass that could be present in neutral, atomic form, located in a photo-dissociation region (PDR) between the inner ion-rich region, and the outer, molecule-rich one. The reason for this is the lack of observations of spectral features suitable for determining low-excitation, neutral masses, i.e. the [C ~\textsc{ii}] 158 \mm{} line, unobservable with ground-based telescopes \citep[see e.g.;][]{castro01,fong01}. Indeed, to our knowledge, the only existing observations of post-CE PNe at this wavelength are unpublished data of NGC~2392 by HERSCHEL/HIFI+PACS, which allow us to estimate that the neutral mass of this nebula amounts to a mere 2$\times$10$^{-3}$ \msun\ (Santander-Garc\'ia et al., in preparation). This is in line with the derived values of the neutral atomic mass in other studied PNe, which is almost always $\lesssim$0.1~\msun (\citealp{castro01}, \citealp{fong01}). This, together with the lack of molecular emission from most post-CE SD PNe and every DD PNe in the sample, hints at the possibility that the gas surrounding these systems tends to be fully ionised. The neutral mass of these systems is unlikely to be substantial enough to change the findings of this paper, although it certainly merits to be the focus of future work (Santander-Garc\'ia et al., in preparation).

\begin{figure*}
\begin{center}
 \includegraphics[width=2\columnwidth]{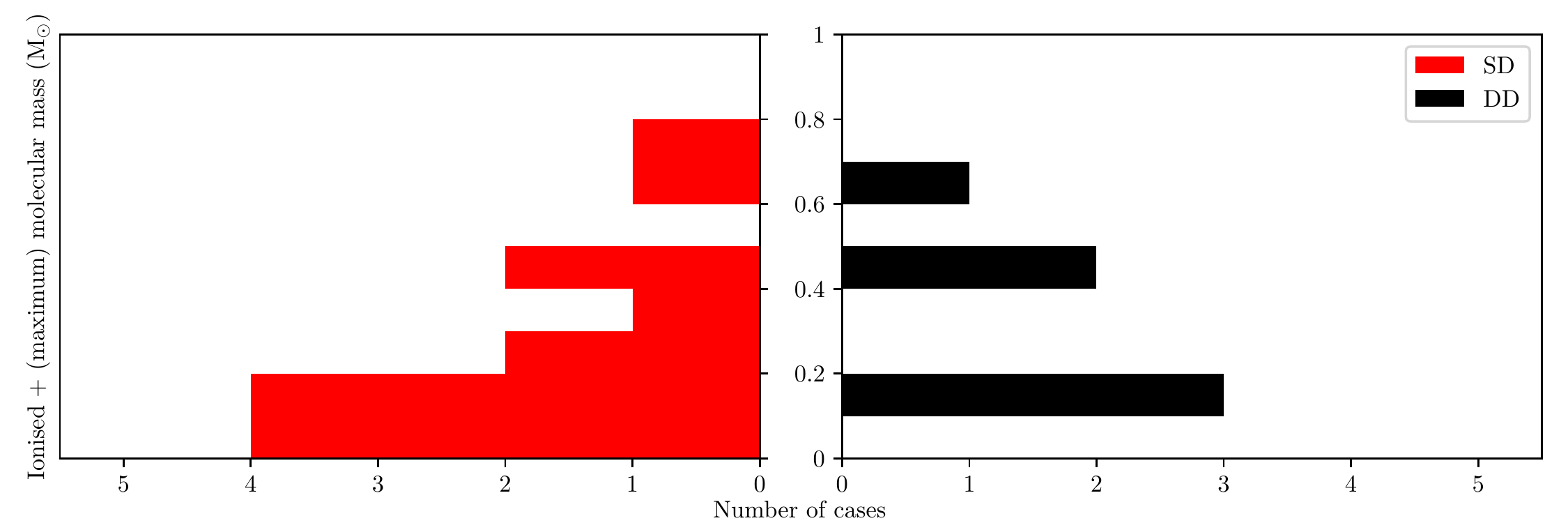}
 \caption{Distribution of the sum of ionised and (maximum) molecular mass of the samples of Single-Degenerate (SD) and Double-Degenerate post-CE PNe analysed in this work.}
   \label{fig:histogram-postce}
\end{center}
\end{figure*}

Fig.~\ref{fig:histogram-postce} shows the mass distribution of each subclass, where a given nebula falls inside a mass bin according to its ionised+molecular mass, or, in the absence of the latter, the sum of its ionised mass and the upper limit to its molecular mass. Hence, the mass distribution plotted there provides a conservative idea of the total ionised+molecular mass the post-CE PNe may have. Even though they fall in the same range, it appears from Fig.~\ref{fig:histogram-postce} and Table~\ref{tab:postcemass} that PN surrounding DD systems tend to be more massive than those around SD binaries. In fact, the geometric mean of the (so-defined) ionised+molecular mass for the SD sample is 0.15~\msun, with a geometric standard deviation (GSD) factor of 3.4, whereas for the DD sample the geometric mean is substantially larger, 0.31~\msun, with a narrower GSD of 1.7.

We can also make an educated guess about the linear momentum and kinetic energy displayed by these objects (again, neglecting any neutral mass that may be present, and treating upper limits to the molecular mass as the molecular mass itself). To this respect, we used characteristic expansion velocities found throughout the literature (prioritising systematic works such as that by \citealp{weinberger89} which take the velocity of the nebula close to the central star and along the line of sight as the characteristic expansion velocity). These can be found, alongside the parameters used in this paper, in Table~\ref{tab:parameters}. We were able to find expansion velocities for every object of the sample except for four SD systems. These seem to follow a similar trend to ionised+molecular mass, being somewhat larger in DD systems than in SD ones.

The resulting linear momenta have substantially different geometric means of 6.3$\times$10$^{38}$ \gcms\ (with GSD factor 3.5) and 2.2$\times$10$^{39}$ \gcms\ (with GSD factor 2.3), for SD and DD systems respectively. As for the kinetic energy of the outflows, their geometric means differ in an even more pronounced way, being 8.1$\times$10$^{44}$~erg (with GSD factor 3.7) for SD systems, and 3.9$\times$10$^{45}$~erg (with GSD factor 4.2) for DD ones. In summary, it seems that both the mass and the velocity (and therefore the linear momentum, and particularly the kinetic energy) of DD post-CE PNe are larger, in general, than those of their SD counterparts.

\subsection{Comparison with regular PNe}\label{comparison}

In this section we try to put previous findings in the context of the general population of PNe. Are post-CE PNe more massive on average than the general population of PNe, as we wondered back in section~\ref{intro}? The answer to this question, as elusive as it may be, may have strong implications for theories of formation of PNe via CE interaction.

In order to bring some insight into this topic, we built an additional, larger sample consisting of `regular' PNe, that is, PNe showing no evidence of hosting a close-binary system. Note that this may include both genuine single-star PNe as well as PNe hosting still undetected post-CE binaries or mergers. A PN had to fulfill the following criteria in order to be included in the regular sample: \textit{i)} being listed in \citealt{frew16}, thus having available a dereddened H$\alpha$ flux and diameters obtained in a systematic way; \textit{ii)} having available \doce\ observations (whether detected or not) to allow accounting for its molecular mass (or upper limit to it); \textit{iii)} having an accurate distance determination, that is, a GAIA eDR3 measurement as identified by \citeauthor{chornay20} (\citeyear{chornay20,chornay21}) with an associated error $<33\%$ or, lacking those, being listed as `distance calibrator' by \cite{frew16} in their table 3; \textit{iv)} having an available determination of its characteristic electronic density $n_e$ (based on the [S~\textsc{ii}] doublet wherever possible).

We therefore built a sample consisting of 97 PNe, essentially by cross-matching the catalog by \cite{frew16} with the molecular surveys by \cite{huggins89}, \cite{huggins96}, \cite{huggins05}, and \cite{guzman18}, rejecting those PNe whose distance was not accurate enough, or for which there was no available measurement of their $n_e$. We prioritised $T_e$ determinations based on [O~\textsc{iii}] where possible, and assumed a $T_e$=10\,000 K wherever no temperature determination was available in a literature search. The only exceptions to this approach are those of NGC~6302, NGC~7027 and NGC~7354, for which we used He-corrected molecular masses found by \citeauthor{santander17} (\citeyear{santander17}, \citeyear{santander12}), and Verbena et al.\ (in preparation), respectively. We highlight that NGC~6302 is the only case in the whole sample analysed in this work whose molecular data includes interferometric measurements subject to flux-loss, but we include it nevertheless, since the analysis by \citeauthor{santander17} found the same mass as in the previous analysis by \citeauthor{santander15a}, which included several singled-dish HERSCHEL/HIFI transitions at frequencies at which the telescope beam FWHM was as large as 20 arcsec, and found interferometric flux-loss to be moderate. The sample is listed in table~\ref{tab:parameters} along with every parameter used in this study.

Note that such a sample is not limited by volume and is thus not exempt from selection biases. Whereas \cite{huggins89} and \cite{huggins96} selected their sample to include objects thought to be at distances shorter than 4~kpc and showing a broad range of properties (morphology, age, abundance, projected size, etc.), and \cite{frew16} made an effort for their sample to be as free of systematic biases as possible, the biases introduced by filtering the intersecting sample by accurate distance determination (and $n_e$ estimates) are difficult to predict. For a truly unbiased sample we would need flux-calibrated [S~\textsc{ii}], [O~\textsc{iii}], H$\alpha$, \doce\ or \trece\ observations of every PNe within a distance sufficiently large for the whole sample to be statistically meaningful, which is clearly out of the scope of this work. In any case, we stress the intrinsic limitation of the comparison provided in this section, which should be taken with a pinch of salt until the wealth of data in the literature is sufficient for this purpose, or until future, ambitious observational efforts to construct such a sample are realised.

\begin{figure}
\begin{center}
 \includegraphics[width=\columnwidth]{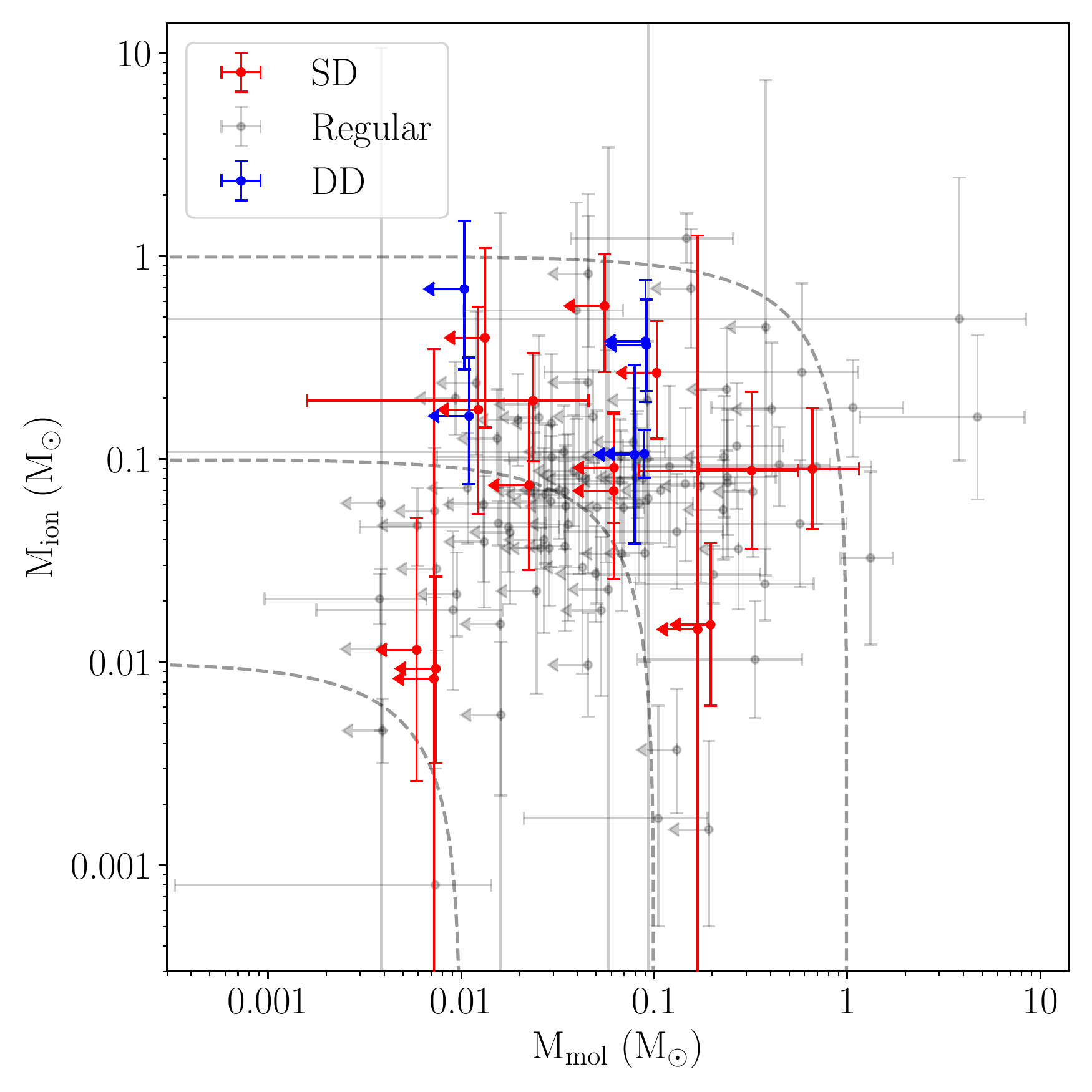}
 \caption{Ionised vs. molecular mass of our post-CE PNe sample and the comparison, regular PNe sample. The further to the top and to the right a nebula is, the more massive it is. Dashed lines represent `isomasses', indicating equal ionised+molecular mass; neglecting neutral atomic mass (see Section~\ref{results}), individual nebulae run along these lines as their gas content is progressively ionised.}
   \label{fig:ionvsmolmass-every}
\end{center}
\end{figure}

\begin{figure}
\begin{center}
 \includegraphics[width=\columnwidth]{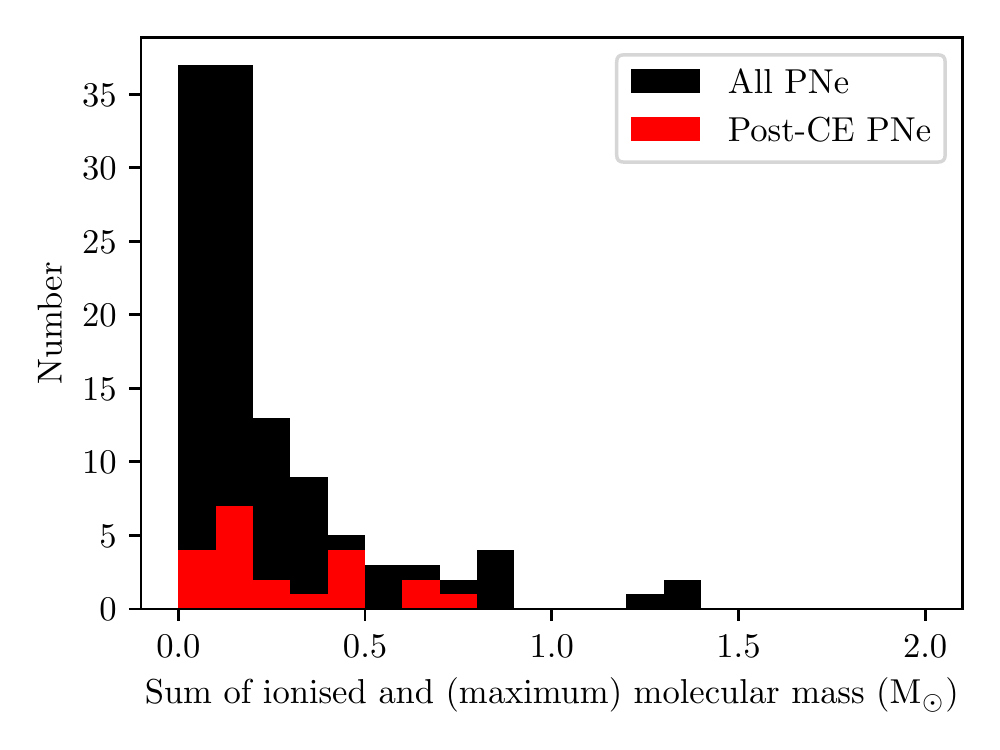}
 \caption{Distribution of the sum of ionised and (maximum) molecular mass of the post-CE PNe sample, along with every PNe (both regular and post-CE PNe) analysed in this work.}
   \label{fig:histogram-every}
\end{center}
\end{figure}

We computed the ionised and molecular masses (or their upper limits) of the whole sample of regular PNe by the same method we followed for estimating the ionised and molecular masses of our sample of post-CE in sections~\ref{ionisedmass-postce} and \ref{molecularmass-postce}, making the same assumptions ($T_\mathrm{exc}$, \doce\ abundance, etc.) where applicable. Results can be found in table~\ref{tab:regularmass}, and are plotted along with the results for the post-CE sample in Figures~\ref{fig:ionvsmolmass-every} and \ref{fig:histogram-every}.

A $k$-sample Anderson-Darling test (\citealp{scholz87}) on the ionised + (maximum) molecular masses of the different samples may provide additional insights. This test is unable to ascertain whether the observed mass distributions for SD and DD systems are different with a probability larger than 75\% (test statistic value = 0.31). The same happens when testing the SD and regular PNe samples (with a test statistic value = 0.055). Despite that, the probability that the whole post-CE sample and the regular PNe sample actually represent different distributions is 80\%, probability that increases to 92\% if we consider only the DD and regular PNe samples.

\begin{figure}
\begin{center}
 \includegraphics[width=\columnwidth]{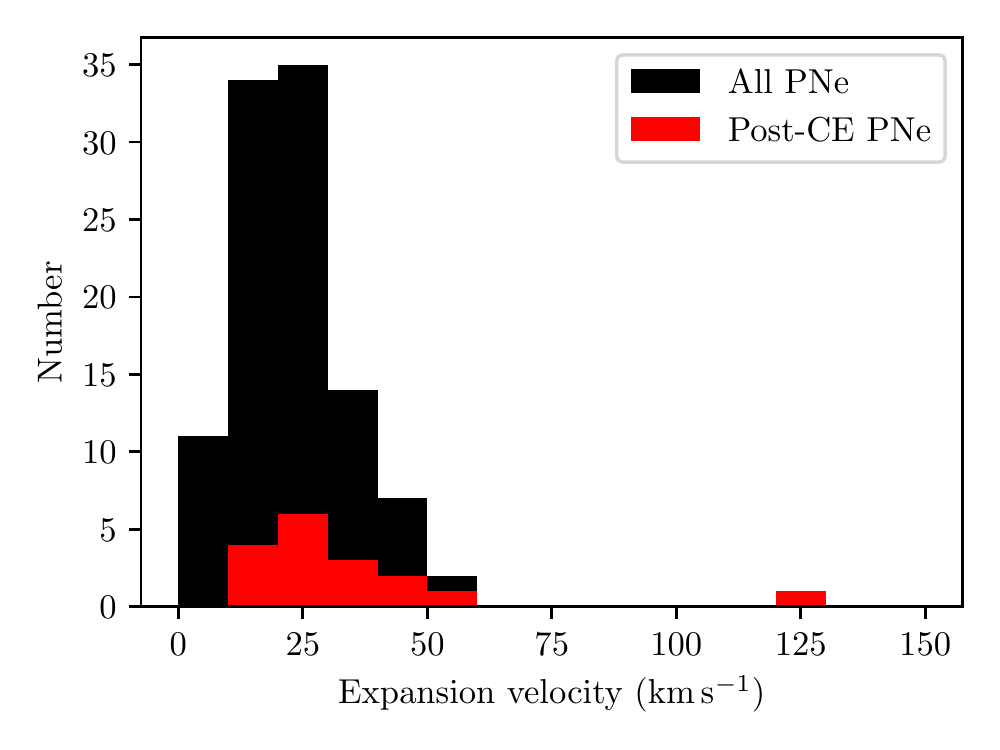}
 \caption{Distribution of the expansion velocity of the post-CE PNe and regular PNe samples analysed in this work.}
   \label{fig:histogram-vexp}
\end{center}
\end{figure}

As before, we also collected the characteristic expansion velocities of (almost) the whole sample, by prioritising systematic works such as that by \cite{weinberger89} wherever possible. On a comparison between the resulting distributions (Figure~\ref{fig:histogram-vexp}), the expansion velocities of the post-CE PNe sample are apparently larger, on average, than those of regular PNe (see Figure~\ref{fig:histogram-vexp}). In fact, the geometric mean of the expansion velocity of the former is 28.9 \kms\ (with GSD factor 1.7), while it is 18.6 \kms\ for regular PNe (with the same GSD factor, 1.7).

While individual values are probably not particularly accurate, the geometric means of the different parameters are likely to be representative of the sample (along with any biases), and therefore allow us to gain some insight on this matter. As such, we used expansion velocities to compute the characteristic linear momentum and kinetic energy of each nebula. The geometric means for the ionised + (maximum) molecular masses, linear momenta and kinetic energies of the different samples can be found in Table~\ref{tab:geometricmeans}.

\begin{table*}
\caption{Geometric means of the ionised + (maximum) molecular mass, linear momentum and kinetic energy of the SD, DD, and regular samples analysed in this work, along with their respectives geometric standard deviation (GSD) factors.}             
\label{tab:geometricmeans}      
\centering                          
\begin{tabular}{l c c c c c c}        
\hline\hline                 
Sample & Mass & GSD$_\mathrm{mass}$ & Momentum &  GSD$_\mathrm{mom}$ & Kinetic Energy & GSD$_\mathrm{kin.energy}$ \\
       & $M_\mathrm{ion}+M_\mathrm{mol}$ (\msun) & & $P$(10$^{38}$ \gcms) & & $E$(10$^{44}$ erg) & \\
\hline\hline
Regular PNe & 0.15 & 3.1 & 5.7 & 3.2 & 5.3 & 4.5 \\
Single-Degenerate post-CE PNe & 0.15 & 3.4 & 6.3 & 3.5 & 8.1 & 3.7 \\
Double-Degenerate post-CE PNe & 0.31 & 1.7 & 22 & 2.3 & 39 & 4.2 \\
\hline
\end{tabular}
\end{table*}

Our results hint towards the following conclusions: the characteristic mass of SD post-CE PNe is indistinguishable from that of the general PNe population. The linear momenta of the SD and regular sample are also very similar, although the slightly larger expansion velocities shown by post-CE systems make the kinetic energy of SD post-CE PNe somewhat larger than that of regular PNe. Meanwhile, the substantially larger masses, as well as the larger expansion velocities found in DD post-CE systems, make their characteristic linear momentum and kinetic energy stand out from the general PN population, and from their SD counterparts. These conclusions hold even when correcting for morphological effects (the fact that post-CE PNe are mostly bipolar) by reducing the expansion velocity of every nebula according to its aspect ratio (see e.g. \citealp{schwarz08}).

\section{Discussion}\label{discussion}

Simple considerations based on the mass lost by a star during the latter phase of the AGB, the briefness of the CE phase and its sudden ejection, would suggest that PNe hosting post-CE systems should be, on average, more massive than PNe arising from single stars (see Section~\ref{intro} and \citealp{boffin19}). Previous work by \cite{frew07} and \citep{corradi15}, however, found the ionised mass of post-CE PNe to be actually lower, on average, than that of the general population of PNe. The analysis presented here considerably expands the sample size to one fifth of the currently known post-CE PNe, and incorporates the molecular content of the nebulae, of which only three systems are detected (including NGC~6778, first reported in this work and analysed in section \ref{ngc6778}). Considered globally, our results suggest a different conclusion: on average, PNe arising from Single-Degenerate (SD) systems seem to be just as massive as `regular' PNe, whereas PNe arising from Double-Degenerate (DD) systems look considerably more massive than both groups.

Differences between samples broaden when considering the linear momentum and kinetic energy of the outflows: since post-CE PNe also show larger expansion velocities (see Figure~\ref{fig:histogram-vexp}), these magnitudes in post-CE PNe depart from the general population of PNe (see Table~\ref{tab:geometricmeans}). This departure is especially notable in the case of DD systems, which are seemingly able to unbind a larger amount of matter than SD systems, and eject it with a larger velocity, thus imprinting their nebulae with an amount of linear momentum and kinetic energy that could help reveal their close-binary origin, should the results of this work be confirmed and generalised by future research. To this respect, it is interesting to note the generally larger masses of the companions in DD systems, mostly over 0.6~\msun, compared to those in SD systems, mostly below 0.4~\msun (Hillwig, private comm.). Perhaps such a difference in companion mass (or the much larger difference in ultraviolet flux) could help explain the observed discrepancy between the nebula ejected by post-CE SD and DD systems.

Our results further suggest a severe mismatch between observations and modelling. As summarised in section~\ref{intro}, models of CE ejection tend to fail in unbinding the whole envelope without the aid of additional energy sources, such as recombination of the ionised region. The observational data instead seem to suggest that the unbound, expanding nebulae do not consist of the whole envelope of their AGB progenitor, but are instead considerably less massive. To this respect, reconstructing the stellar and orbital parameters at CE onset in these systems may help assessing the fraction of the AGB envelope actually ejected, as well as the fraction of the orbital energy budget (the change in energy from orbital shrinkage) spent on unbinding and accelerating the nebula to the observed expansion velocity. While such an effort is undoubtedly plagued with caveats and large uncertainties, it can provide an `order of magnitude estimate' to help guide theoretical modelling efforts.

Following the methodology described in \cite{iaconi19} and \cite{demarco11}, we have attempted reconstruction of the CE of the two SD systems, Abell~63 and Hen~2-155, and the two DD systems, Fg~1 and Hen~2-428, for which we have sufficient information on the orbital parameters. Table~\ref{tab:cereconstruction} shows the calculated efficiency $\alpha$, the estimated AGB envelope mass of the primary star, $M_\mathrm{env}$, the percentage of the envelope mass contained in the observed (ionised+molecular) nebula, $f_M$, the orbital energy budget, $\Delta E_\mathrm{orb}$, a rough estimate of the binding energy of the observed nebula with respect to the primary star core (assuming the $\lambda$ parameter for AGBs provided by \citealp{demarco11}), its kinetic energy, and the percentage of the energy budget spent on unbinding and accelerating the nebula, $f_E$, along with the references for the orbital parameters used.

\begin{table*}
\caption{CE and nebula ejection reconstruction parameters for a sample of post-CE PNe. Columns represent CE ejection (whole envelope) efficiency $\alpha$, AGB envelope mass $M_\mathrm{env}$, percentage of envelope mass each nebula represents, $f_M$, orbital energy budget, $\Delta E_\mathrm{orb}$, rough estimate of the observed nebula binding energy, $E_\mathrm{bin,neb}$, observed nebula kinetic energy, $E_\mathrm{kin,neb}$, and percentage of the energy budget spent on unbinding and accelerating the nebula, $f_E$.}             
\label{tab:cereconstruction}      
\centering                          
\begin{tabular}{l c c c c c c c l}        
\hline\hline                 
Nebula & $\alpha$ & $M_\mathrm{env}$ & $f_M$ & $\Delta E_\mathrm{orb}$ & $E_\mathrm{bin,neb}$ & $E_\mathrm{kin,neb}$ &  $f_E$ & References \\
       &  & (\msun) &  & (erg) & (erg) & (erg)  &  & \\
\hline\hline
\multicolumn{9}{c}{\sc{Single-Degenerate post-CE PNe}}\\
\hline
Abell~63 & 0.29 & 1.7 & 1\% & 1.4$\times$10$^{47}$ & 8.3$\times$10$^{44}$ & 5$\times$10$^{43}$ & 0.6\% & \cite{demarco08}, \cite{iaconi19} \\
Hen~2-155 & 0.3 & 1.7 & 22\% & 1.4$\times$10$^{47}$ & 9.7$\times$10$^{45}$ & 2.9$\times$10$^{45}$ & 9\%$^a$ & \cite{jones15}, \cite{iaconi19} \\
\hline\hline
\multicolumn{9}{c}{\sc{Double-Degenerate post-CE PNe}}\\
\hline
Fg~1 & 0.11$^b$ & 0.7 & 64\% & 1.4$\times$10$^{47^b}$ & 1$\times$10$^{46}$ & 5.9$\times$10$^{45}$ &  11\% & \cite{boffin12} \\
Hen~2-428 & 0.04 & 1.1 & 61\% & 4$\times$10$^{47}$ & 9.8$\times$10$^{45}$ & 1.6$\times$10$^{45}$ & 3\% & \cite{reindl20}\\
\hline
\end{tabular}
\tablefoot{
\tablefoottext{a}{Characteristic expansion velocity unavailable, average of expansion velocities in SD systems used (see Table~\ref{tab:parameters}).}
\tablefoottext{b}{Mass of the secondary star (in the range 0.7-1 \msun\ according to radial velocity analysis, and in the range 0.63-0.7 \msun\ according to evolutionary and ionisation considerations), assumed to be 0.7~\msun.}
}
\end{table*}

If confirmed and generalised by additional data on the orbital parameters of other systems, these results seem to suggest that post-CE PNe arising from SD systems are substantially less massive than the envelope of their AGB progenitors, while those arising from DD systems comprise of most (if not all, given such large uncertainties) of their progenitors' envelopes.

In any case, a problem akin to the long standing issue of the missing mass of PNe (e.g.~\citealp{kimura12}) persists. While one could in principle think that the mass we do not detect in `regular' PNe is long gone, diluted in the ISM after millions of years of AGB wind in, the fact that we cannot reconcile the observed mass of SD post-CE PNe with the mass of their envelopes at the time of CE interaction should constitute a warning call about our incomplete understanding of the physics behind CE ejection. The missing mass in SD systems thus leads to uncomfortable questions: if the primary star is of similar mass to normal post-AGB stars, and thus the mass of the nebula amounts to just a tiny fraction of the star's envelope, then, where is the rest of the envelope? Why are we unable to detect it somewhere in the star's vicinity?

From a theoretical perspective, we can consider some possibilities. A fraction of the ejected mass could fall back and form a circumbinary disk (as in \citealp{kuruwita16}). If any of this material reaches the central stars, it could then be reprocessed, which could offer an explanation for the correlation between large abundance discrepancy factors and post-CE central stars in PNe \cite{wesson18}. We can thus wonder whether the CE itself could be not a unique, once-only process, but a long lasting or episodic one. Models such as the Grazing Envelope Evolution proposed by \cite{soker15} and \cite{shiber17}, in which the companion grazes the envelope of the RGB or AGB star while both the orbital separation and the giant radius shrink simultaneously over the course of tens to hundreds of years, could perhaps help explain the phenomenon. A model along these lines could, for instance, provide some insight in the case of NGC~2346, a particularly massive SD system with a relatively long post-CE orbital period in which the primary is believed to be a post-RGB star (\citealp{brown19}). At any rate, CE interaction would probably need to last long enough to allow for a considerable amount of the primary's envelope mass to get diluted into the ISM beyond detectability, in order for the presumed envelope mass at the time of a later ``full'' CE ejection to be reconciled with the mass of the observed nebulae.

From an observational point of view, instead, it can be also interesting to study the infrared grain emission and the mass in dust in these objects. There exists the possibility that some amount of mass is contained in low-excitation neutral atoms, mainly in a PDR between the ionised region and an outer, molecule-rich domain. This is however unlikely, given the general lack of molecular content observed in post-CE PNe reported in this work, which would suggest most of these nebulae are (almost) fully ionised. In fact, the only available data on a post-CE PN, a set of HERSCHEL/HIFI+PACS observations of NGC~2392, points to a very low neutral mass, of the order of 2$\times$10$^{-3}$ \msun\ (Santander-Garc\'ia et al., in preparation). In spite of this, assessing the amount of low-excitation, neutral mass of a sample of post-CE PNe is still missing, and will be the subject of future work by our group.

\section{Conclusions}\label{conclusions}

In this work we have gathered literature-available data on post-CE PNe including dereddened H$\beta$ fluxes, and carried out observations of the molecular content of these objects, totalling a sample of 21 objects, roughly one fifth of the total known population of post-CE PNe. We find a general lack of molecular content, with the exceptions of NGC~2346 and NGC~7293 in the literature, and our observations of NGC~6778. The data on the latter have allowed us to study the physical conditions of the molecular gas, as well as its spatial distribution, with the CO-rich gas located in a ring lying beyond the broken, clumpy ionised equatorial ring described by \cite{guerrero12}, and expanding alongside it.

By means of the systematic calculation of the ionised and molecular masses of the whole sample of post-CE PNe, we conclude that post-CE PNe with SD central stars are as massive, on average, as their single star counterparts, whereas post-CE PNe with DD central stars are considerably more massive than both groups. The characteristic expansion velocities of post-CE PNe also seem larger than those of regular PNe. This in turn results in larger linear momenta and kinetic energy of the ejecta, which are particularly notable in the case of DD post-CE PNe.

We have reconstructed the CE in the four systems (two SD and two DD) for which sufficient data on the orbital parameters are available, including the present masses of both stars, the orbital separation, and the presumed envelope mass of the primary at the time of CE. We find that DD systems eject more massive nebulae at larger velocities than SD systems do. We cannot, however, reconcile the observed mass of the nebulae with the presumed mass of the progenitor star envelopes. Whereas PNe around DD systems would contain most (if not all) of the envelope of the progenitor AGB star, PNe in SD systems (as well as in 'regular' PNe) only show but a small fraction of the progenitor envelope. This in turn leads to an alarming question: if the remaining mass of the envelope of these systems is no longer on the surface of the now post-AGB star, and is not contained in the ejected CE (the nebula), where is it? The possibility that the large amount of missing mass in SD systems is in a hard-to-detect halo beyond the PN would in principle require CE interaction to last much longer than commonly found by models, in order for the AGB star to be able to dispose of most of its envelope beyond detectability before shaping the visible nebula. Similarly, although the models of \citet{vigna-gomez21} find that an appreciable quantity of the envelope may remain on the surface of the CE donor (the central star) following ejection, the discrepancies between observed and expected post-CE PN masses are too large to be explained by this scenario alone.

Future efforts to answer this question will in any case require new theoretical work on the one hand, and systematic observations of the ionised and molecular content of the whole known population of post-CE PNe on the other, as well as observationally assessing the (unlikely) possibility that a significant fraction of these nebulae consists of low-excitation, neutral gas yet to be studied.

%

\begin{acknowledgements}
MSG, JA, and VB acknowledge  support by the Spanish Ministry of Science and Innovation (MICINN) through projects AxIN (grant AYA2016-78994-P) and EVENTs/Nebulae-Web (grant PID2019-105203GB-C21 ). DJ acknowledges support from the Erasmus+ programme of the European Union under grant number 2020-1-CZ01-KA203-078200. DJ also acknowledges support under grant P/308614 financed by funds transferred from the Spanish Ministry of Science, Innovation and Universities, charged to the General State Budgets and with funds transferred from the General Budgets of the Autonomous Community of the Canary Islands by the Ministry of Economy, Industry, Trade and Knowledge.

This work has made use of data from the European Space Agency (ESA) mission
{\it Gaia} (\url{https://www.cosmos.esa.int/gaia}), processed by the {\it Gaia} Data Processing and Analysis Consortium (DPAC, \url{https://www.cosmos.esa.int/web/gaia/dpac/consortium}). Funding for the DPAC has been provided by national institutions, in particular the institutions participating in the {\it Gaia} Multilateral Agreement.
\end{acknowledgements}


\bibliographystyle{aa}
\bibliography{aa-2021-42233}

\begin{appendix}
\section{The mass of the regular comparison sample}

Table~\ref{tab:regularmass} shows the computed ionised and molecular masses of the sample of 97 regular PNe used for comparison in section~\ref{comparison} the main paper.

\onecolumn
\begin{longtable}{llccc}
\caption{\label{tab:regularmass} Computed ionised and molecular masses of the regular comparison sample. Masses as determined here scale with distance squared.} \\
\hline\hline                 
PN G & Common name & $D$ & M$_\mathrm{ion}$ & M$_\mathrm{mol}$ \\
    &             &  (kpc)   &    (\msun)         & (\msun)  \\
\hline
\endfirsthead
\caption{continued.}\\
\hline\hline
PN G & Common name & $D$ & M$_\mathrm{ion}$ & M$_\mathrm{mol}$ \\
    &             &  (kpc)   &    (\msun)         & (\msun)  \\
\hline
\endhead
\hline
\endfoot
G001.5-06.7 & SwSt~1 & 2.941$\pm$0.952 & 0.02$^{+0.04}_{-0.015}$ & $<$0.02 \\
G002.4+05.8 & NGC~6369 & 1.087$\pm$0.059 & 0.16$^{+0.11}_{-0.06}$ & $<$0.020 \\
G002.8+01.7 & H~2-20 & 8.3$\pm$2.4 & 0.03$^{+0.06}_{-0.02}$ & $<$0.04 \\
G008.3-07.3 & NGC~6644 & 8.3$\pm$2.4 & 0.12$^{+0.10}_{-0.06}$ & $<$0.08 \\
G009.4-05.0 & NGC~6629 & 2.041$\pm$0.083 & 0.19$^{+0.14}_{-0.08}$ & $<$0.02 \\
G009.6+14.8 & NGC~6309 & 2.632$\pm$0.416 & 0.07$^{+0.06}_{-0.03}$ & $<$0.03 \\
G010.7-06.4 & IC~4732 & 8.3$\pm$2.4 & 0.05$^{+0.09}_{-0.03}$ & $<$0.04 \\
G011.9+04.2 & M~1-32 & 2.632$\pm$0.416 & 0.018$^{+0.03}_{-0.011}$ & 0.009$\pm$0.007 \\
G016.4-01.9 & M~1-46 & 2.381$\pm$0.113 & 0.06$^{+0.08}_{-0.03}$ & $<$0.013 \\
G021.8-00.4 & M~3-28 & 2.5$^{+1.1}_{-1.3}$ & 0.5$^{+1.9}_{-0.4}$ & 3.8$\pm$4.6 \\
G025.3+40.8 & IC~4593 & 2.632$\pm$0.346 & 0.07$^{+0.06}_{-0.03}$ & $<$0.11 \\
G033.1-06.3 & NGC~6772 & 0.901$\pm$0.146 & 0.10$^{+0.17}_{-0.06}$ & 0.09$\pm$0.07 \\
G033.8-02.6 & NGC~6741 & 2.6$\pm$0.55 & 0.15$^{+0.18}_{-0.08}$ & $<$0.03 \\
G034.6+11.8 & NGC~6572 & 1.852$\pm$0.206 & 0.04$^{+0.06}_{-0.02}$ & $<$0.03 \\
G035.9-01.1 & Sh~2-71 & 1.613$\pm$0.052 & 0.18$^{+0.20}_{-0.09}$ & $<$0.4 \\
G037.7-34.5 & NGC~7009 & 1.235$\pm$0.091 & 0.09$^{+0.08}_{-0.04}$ & $<$0.04 \\
G038.2+12.0 & Cn~3-1 & 7.143$\pm$1.531 & 0.08$^{+0.17}_{-0.06}$ & $<$0.04 \\
G041.8-02.9 & NGC~6781 & 0.500$\pm$0.018 & 0.05$^{+0.05}_{-0.02}$ & 0.6$\pm$0.4 \\
G043.1+03.8 & M~1-65 & 6.667$\pm$0.889 & 0.08$^{+0.08}_{-0.04}$ & $<$0.07 \\
G043.1+37.7 & NGC~6210 & 2.041$\pm$0.125 & 0.07$^{+0.06}_{-0.03}$ & $<$0.011 \\
G045.4-02.7 & Vy~2-2 & 3.5$\pm$1.2 & 0.11$^{+0.2}_{-0.08}$ & 0.06$\pm$0.06 \\
G051.4+09.6 & Hu~2-1 & 2.381$\pm$0.397 & 0.05$^{+0.03}_{-0.017}$ & $<$0.006 \\
G051.9-03.8 & M~1-73 & 4.545$\pm$0.620 & 0.04$^{+0.05}_{-0.02}$ & $<$0.018 \\
G052.5-02.9 & Me~1-1 & 3.704$\pm$0.274 & 0.02$^{+0.013}_{-0.008}$ & $<$0.009 \\
G055.5-00.5 & M~1-71 & 2.9$\pm$0.4 & 0.2$^{+0.3}_{-0.13}$ & $<$0.012 \\
G055.6+02.1 & Hen~1-2 & 10.000$\pm$3.000 & 0.3$^{+0.5}_{-0.18}$ & 0.6$\pm$0.6 \\
G056.0+02.0 & K~3-35 & 3.9$^{+0.7}_{-0.5}$ & 0.0017$^{+0.004}_{-0.0012}$ & 0.11$\pm$0.08 \\
G058.6+06.1 & Abell~57 & 2.128$\pm$0.317 & 0.04$^{+0.07}_{-0.03}$ & $<$0.03 \\
G060.8-03.6 & NGC~6853 & 0.389$\pm$0.006 & 0.5$^{+1.3}_{-0.4}$ & 0.04$\pm$0.03 \\
G063.1+13.9 & NGC~6720 & 0.787$\pm$0.025 & 0.12$^{+0.12}_{-0.06}$ & 0.3$\pm$0.20 \\
G064.6+48.2 & NGC~6058 & 2.778$\pm$0.231 & 0.010$^{+0.008}_{-0.004}$ & $<$0.05 \\
G064.7+05.0 & BD+303639 & 1.613$\pm$0.078 & 0.05$^{+0.014}_{-0.011}$ & 0.016$\pm$0.012 \\
G068.3-02.7 & Hen~2-459 & 1.010$\pm$0.306 & 0.0008$^{+0.002}_{-0.0006}$ & 0.007$\pm$0.007 \\
G069.4-02.6 & NGC~6894 & 1.449$\pm$0.231 & 0.0015$^{+0.003}_{-0.0010}$ & $<$0.19 \\
G071.6-02.3 & M~3-35 & 1.000$\pm$0.310 & 0.006$^{+0.007}_{-0.003}$ & $<$0.016 \\
G082.1+07.0 & NGC~6884 & 3.3$\pm$1.24 & 0.06$^{+0.06}_{-0.03}$ & $<$0.007 \\
G083.5+12.7 & NGC~6826 & 1.299$\pm$0.067 & 0.04$^{+0.03}_{-0.017}$ & $<$0.03 \\
G084.9-03.4 & NGC~7027 & 0.92$\pm$0.1 & 0.03$^{+0.05}_{-0.02}$ & 1.3$\pm$0.4 \\
G088.7-01.6 & NGC~7048 & 1.587$\pm$0.529 & 0.018$^{+0.02}_{-0.011}$ & $<$0.05 \\
G089.0+00.3 & NGC~7026 & 3.226$\pm$0.312 & 0.10$^{+0.10}_{-0.05}$ & $<$0.2 \\
G093.4+05.4 & NGC~7008 & 0.645$\pm$0.033 & 0.02$^{+0.007}_{-0.005}$ & 0.004$\pm$0.003 \\
G093.9-00.1 & IRAS~21282 & 3.704$\pm$0.274 & 0.16$^{+0.2}_{-0.10}$ & 4.7$\pm$3.6 \\
G096.4+29.9 & NGC~6543 & 1.370$\pm$0.056 & 0.06$^{+0.06}_{-0.03}$ & $<$0.03 \\
G104.2-29.6 & Jn~1 & 0.990$\pm$0.069 & 0.07$^{+0.14}_{-0.05}$ & $<$0.17 \\
G104.4-01.6 & M~2-53 & 6.0$\pm$1.0 & 0.18$^{+0.13}_{-0.08}$ & 1.1$\pm$0.9 \\
G106.5-17.6 & NGC~7662 & 1.754$\pm$0.092 & 0.13$^{+0.09}_{-0.05}$ & $<$0.015 \\
G107.6-13.3 & Vy~2-3 & 6.250$\pm$1.172 & 0.05$^{+0.03}_{-0.018}$ & 0.018$\pm$0.015 \\
G107.8+02.3 & NGC~7354 & 2.083$\pm$0.304 & 0.08$^{+0.10}_{-0.04}$ & 0.2$\pm$0.08 \\
G116.2+08.5 & M~2-55 & 0.658$\pm$0.022 & 0.005$^{+0.002}_{-0.0014}$ & $<$0.004 \\
G118.0-08.6 & Vy~1-1 & 5.000$\pm$0.750 & 0.04$^{+0.03}_{-0.018}$ & $<$0.3 \\
G120.0+09.8 & NGC~40 & 1.786$\pm$0.064 & 0.11$^{+0.07}_{-0.04}$ & $<$0.03 \\
G123.6+34.5 & IC~3568 & 2.273$\pm$0.207 & 0.06$^{+0.04}_{-0.02}$ & $<$0.07 \\
G130.2+01.3 & IC~1747 & 3.846$\pm$0.592 & 0.08$^{+0.11}_{-0.05}$ & $<$0.04 \\
G138.8+02.8 & IC~289 & 1.587$\pm$0.126 & 0.07$^{+0.08}_{-0.04}$ & $<$0.3 \\
G144.5+06.5 & NGC~1501 & 1.724$\pm$0.059 & 0.2$^{+0.2}_{-0.11}$ & $<$0.2 \\
G146.7+07.6 & M~4-18 & 6.667$\pm$0.889 & 0.02$^{+3.4}_{-0.02}$ & $<$0.06 \\
G147.4-02.3 & M~1-4 & 3.3$\pm$0.35 & 0.07$^{+0.08}_{-0.04}$ & $<$0.03 \\
G148.4+57.0 & NGC~3587 & 0.813$\pm$0.033 & 0.10$^{+0.04}_{-0.03}$ & $<$0.07 \\
G164.8+31.1 & JnEr~1 & 0.943$\pm$0.071 & 0.06$^{+0.12}_{-0.04}$ & 0.05$\pm$0.04 \\
G166.1+10.4 & IC~2149 & 1.852$\pm$0.137 & 0.04$^{+0.04}_{-0.02}$ & $<$0.013 \\
G189.1+19.8 & NGC~2371 & 1.724$\pm$0.149 & 0.16$^{+0.2}_{-0.10}$ & $<$0.03 \\
G193.6-09.5 & H~3-75 & 4.000$\pm$0.320 & 0.4$^{+6.9}_{-0.4}$ & $<$0.4 \\
G194.2+02.5 & J~900 & 4.55$\pm$0.25 & 0.10$^{+0.05}_{-0.03}$ & 0.03$\pm$0.02 \\
G196.6-10.9 & NGC~2022 & 2.273$\pm$0.258 & 0.06$^{+0.04}_{-0.02}$ & $<$0.2 \\
G204.8-03.5 & K~3-72 & 4.6$\pm$0.8 & 0.08$^{+0.09}_{-0.04}$ & $<$0.08 \\
G205.1+14.2 & Abell~21 & 0.592$\pm$0.025 & 0.08$^{+0.03}_{-0.02}$ & $<$0.09 \\
G206.4-40.5 & NGC~1535 & 1.370$\pm$0.113 & 0.03$^{+0.020}_{-0.012}$ & $<$0.05 \\
G211.2-03.5 & M~1-6 & 7.692$\pm$2.367 & 0.06$^{+16.}_{-0.06}$ & $<$0.09 \\
G215.2-24.2 & IC~418 & 1.370$\pm$0.056 & 0.06$^{+11.}_{-0.06}$ & $<$0.004 \\
G217.1+14.7 & Abell~24 & 0.719$\pm$0.052 & 0.2$^{+1.8}_{-0.2}$ & $<$0.05 \\
G221.7+05.3 & M~3-3 & 5.5$^{+1.3}_{-1.8}$ & 0.09$^{+0.08}_{-0.04}$ & 0.7$\pm$0.6 \\
G226.4-03.7 & PB~1 & 3.226$\pm$0.937 & 0.015$^{+1.6}_{-0.015}$ & $<$0.016 \\
G231.8+04.1 & NGC~2438 & 0.725$\pm$0.116 & 0.03$^{+0.04}_{-0.017}$ & $<$0.007 \\
G232.8-04.7 & M~1-11 & 4.545$\pm$0.620 & 0.07$^{+0.07}_{-0.03}$ & $<$0.02 \\
G234.8+02.4 & NGC~2440 & 1.77$\pm$0.45 & 0.06$^{+0.08}_{-0.04}$ & 0.08$\pm$0.08 \\
G235.3-03.9 & M~1-12 & 4.545$\pm$0.620 & 0.03$^{+0.03}_{-0.016}$ & $<$0.07 \\
G243.3-01.0 & NGC~2452 & 2.941$\pm$0.692 & 0.07$^{+0.04}_{-0.03}$ & $<$0.03 \\
G261.0+32.0 & NGC~3242 & 1.333$\pm$0.089 & 0.16$^{+0.11}_{-0.07}$ & $<$0.05 \\
G261.9+08.5 & NGC~2818 & 3.0$\pm$0.8 & 0.10$^{+0.07}_{-0.05}$ & 0.08$\pm$0.07 \\
G272.1+12.3 & NGC~3132 & 0.758$\pm$0.017 & 0.04$^{+0.04}_{-0.02}$ & 0.13$\pm$0.10 \\
G277.1-03.8 & NGC~2899 & 1.923$\pm$0.111 & 1.2$^{+0.4}_{-0.3}$ & 0.15$\pm$0.11 \\
G279.6-03.1 & Hen~2-36 & 4.000$\pm$0.160 & 0.7$^{+0.7}_{-0.3}$ & $<$0.16 \\
G283.8-04.2 & Hen~2-39 & 7.6$^{+1.5}_{-1.3}$ & 0.19$^{+0.19}_{-0.09}$ & $<$0.09 \\
G292.6+01.2 & NGC~3699 & 1.370$\pm$0.150 & 0.06$^{+0.07}_{-0.03}$ & $<$0.03 \\
G294.1+43.6 & NGC~4361 & 1.031$\pm$0.043 & 0.04$^{+0.010}_{-0.008}$ & $<$0.03 \\
G294.6+04.7 & NGC~3918 & 4.545$\pm$1.446 & 0.8$^{+0.8}_{-0.5}$ & $<$0.05 \\
G294.9-04.3 & Hen~2-68 & 7.692$\pm$2.367 & 0.03$^{+0.08}_{-0.02}$ & $<$0.09 \\
G309.1-04.3 & NGC~5315 & 0.962$\pm$0.185 & 0.012$^{+0.017}_{-0.007}$ & $<$0.004 \\
G315.1-13.0 & Hen~2-131 & 2.703$\pm$0.219 & 0.20$^{+0.10}_{-0.07}$ & $<$0.009 \\
G319.6+15.7 & IC~4406 & 1.136$\pm$0.155 & 0.02$^{+0.012}_{-0.008}$ & 0.4$\pm$0.3 \\
G321.0+03.9 & Hen~2-113 & 2.083$\pm$0.130 & 0.03$^{+0.011}_{-0.008}$ & 0.2$\pm$0.15 \\
G322.4-02.6 & Mz~1 & 1.266$\pm$0.208 & 0.08$^{+0.10}_{-0.04}$ & 0.14$\pm$0.12 \\
G326.7+42.2 & IC~972 & 2.222$\pm$0.494 & 0.004$^{+0.004}_{-0.0019}$ & $<$0.13 \\
G332.9-09.9 & Hen~3-1333 & 1.471$\pm$0.108 & 0.010$^{+0.010}_{-0.005}$ & 0.3$\pm$0.3 \\
G342.1+10.8 & NGC~6072 & 0.917$\pm$0.168 & 0.09$^{+0.05}_{-0.04}$ & 0.4$\pm$0.4 \\
G349.5+01.0 & NGC~6302 & 1.17$\pm$0.14 & 0.09$^{+0.14}_{-0.06}$ & 0.12$\pm$0.04 \\
G358.5-07.3 & NGC~6563 & 0.935$\pm$0.114 & 0.08$^{+0.07}_{-0.04}$ & 0.2$\pm$0.19 \\
\hline
\end{longtable}

\section{Parameters used in the analysis}

Table~\ref{tab:parameters} shows the parameters used in this work for computing the ionised and molecular masses, as well as the linear momenta and kinetic energy, of the studied sample of post-CE PNe, as well as the regular PNe comparison sample.

\begin{landscape}
\begin{longtable}{p{2cm}p{1.5cm}p{1.5cm}p{2cm}p{1.5cm}p{1cm}p{1.8cm}p{4.5cm}p{5.5cm}}
\caption{\label{tab:parameters} Electron densities, temperatures, distances, sizes (major axes), expansion velocities, H$\alpha$ fluxes, and \doce\ emission (rms if undetected, Intensity $I$ or spatially-integrated Flux $F$ if detected) used in the analysis.}\\
\hline\hline
PN& $n_e$ & $T_e$ &  $D$  & Diameters & \vexp & log $S_0$(H$\alpha$)  & Tel., \doce $J$ transition, emission & References \\
& (cm$^{-3}$) & (kK) & (kpc) & ($''$) & (\kms) & (cgs sr$^{-1}$) &  &   \\
\hline
\endfirsthead
\caption{continued.}\\
\hline\hline
PN& $n_e$ & $T_e$ & $D$ & Diameters & \vexp & log $S_0$(H$\alpha$) & Telescope, \doce emission & References \\
& (cm$^{-3}$) & (kK) & (kpc) & ($''$) & (\kms) & (cgs sr$^{-1}$) &  &  \\
\hline
\endhead
\hline
\endfoot
\hline\hline
\multicolumn{9}{c}{\textsc{Single-Degenerate post-CE PNe}}\\
\hline
Abell~30 & 200 & 13.6 & 2.222$\pm$0.148 & 127$\times$127 & 40.0 & -5.25$\pm$0.06 & N, \du, rms: 47~mK & $n_e$,$T_e$: 1; $D$: 2; v: 3; CO: 4\\
Abell~63 & 600$^{+2100}_{-450}$ & 7.40$\pm$0.55 & 2.703$\pm$0.219 & 48$\times$42 & 17.0 & -3.93$\pm$0.14 & I, \du, rms: 9.0~mK & $n_e$,$T_e$: 5; $D$: 2; v: 6; CO: This work\\
Necklace & 360$^{+380}_{-240}$ & 14.80$^{+0.53}_{-0.46}$ & 4.6$\pm$1.1 & 13.0$\times$6.7\tablefootmark{b} & 28.0 & -3.59$\pm$0.09\tablefootmark{b} & I, \du, rms: 8.5~mK & $n_e$,$T_e$,$D$,v: 7; CO: This work\\
ETHOS~1 & 850$\pm$1000 & 17.70$\pm$0.50 & 4.2$\pm$0.0 & 19.5$\times$19.0 & 55.0 & -3.89$\pm$0.05 & I, \du, rms: 7.2~mK & $n_e$,$T_e$,v: 8; $D$: 9; CO: This work\\
Hen~2-155 & 1390$\pm$55 & 11.66$\pm$0.04 & 4.348$\pm$1.323 & 18$\times$16 & -- & -1.94$\pm$0.10 & S, \du, rms: 42~mK & $n_e$,$T_e$: 10; $D$: 2; CO: 11\\
Lo~16 & 200$\pm$150 & 11.60$\pm$0.80 & 1.818$\pm$0.132 & 88$\times$80 & -- & -3.24$\pm$0.12 & A, \td, rms: 69~mK & $n_e$,$T_e$: 12; $D$: 2; CO: 13\\
MyCn~18 & 5025 & 7.3 & 4.000$\pm$1.280 & 17.3$\times$9.8 & 24.0 & -1.47$\pm$0.08 & A, \td, rms: 198~mK & $n_e$,$T_e$: 14; $D$: 2; v: 6; CO: 13\\
NGC~246 & 160 & 15.8 & 0.556$\pm$0.025 & 260$\times$227 & 39.0 & -4.08$\pm$0.05 & N, \du, rms: 62~mK & $n_e$: 15; $T_e$: 16; $D$: 2; v: 3; CO: 4\\
NGC~2346 & 265$\pm$60 & 11.6\tablefootmark{d} & 1.389$\pm$0.039 & 124$\times$59 & 28.0 & -3.55$\pm$0.28 & I, \du, $F$: 6.0$\times$10$^{4}$~\kkmsa & $n_e$: 12; $D$: 2; v: 17; CO: 11\\
NGC~6326 & 750$\pm$30 & 14.60$\pm$0.10 & 5.000$\pm$1.500 & 20.6$\times$13.7 & 16.5 & -2.08$\pm$0.11 & A, \td, rms: 90~mK & $n_e$,$T_e$: 12; $D$: 2; v: 3; CO: 13\\
NGC~6778 & 590$\pm$40 & 8.80$\pm$0.08 & 2.79$\pm$0.79 & 21.4$\times$15.5 & 26.0 & -2.02$\pm$0.08 & I, \du, $I$: 1.4~\kkms & $n_e$,$T_e$: 18; $D$: 19; v: 6; CO: This work\\
NGC~7293 & 220 & 11.70$\pm$0.70 & 0.200$\pm$0.002 & 970$\times$735 & 14.0 & -3.95$\pm$0.06 & N, \du, $F$: 1.4$\times$10$^{6}$~\kkmsa & $n_e$: 20; $T_e$: 21; $D$: 2; v: 3; CO: 11\\
Ou~5 & 150$^{+300}_{-100}$ & 10.15$\pm$0.30 & 5.0$\pm$1.0 & 16$\times$14 & -- & -3.04$\pm$0.04 & I, \du, rms: 10~mK & $n_e$,$T_e$: 5; $D$: 22; CO: This work\\
PM~1-23 & 2250$^{+4000}_{-2250}$ & 10\tablefootmark{a} & 5.2$\pm$2.0 & 27$\times$16 & -- & -3.21$\pm$0.38 & I, \du, rms: 100~mK & $n_e$,$T_e$: 12; $D$: 19; CO: This work\\
Sp~3 & 640$^{+270}_{-210}$ & 7.24$\pm$0.15 & 2.22$^{+0.61}_{-0.48}$ & 36$\times$35 & 22.0 & -2.63$\pm$0.07 & S, \du, rms: 52~mK & $n_e$,$T_e$: 23; $D$: 19; v: 3; CO: 11\\
\hline
\multicolumn{9}{c}{\textsc{Double-Degenerate post-CE PNe}}\\
\hline
Abell~41 & 300$\pm$100 & 10\tablefootmark{a} & 4.89$\pm$1.4 & 20.2$\times$17.3 & 40.0 & -2.94$\pm$0.09 & I, \du, rms: 8.2~mK & $n_e$: 12; $D$: 19; v: 24; CO: This work\\
Fg~1 & 290$^{+150}_{-120}$ & 11.00$\pm$0.30 & 2.564$\pm$0.197 & 55$\times$40 & 36.0 & -2.89$\pm$0.06 & S, \du, rms: 48~mK & $n_e$,$T_e$: 12; $D$: 2; v: 6; CO: 11\\
Hen~2-428 & 580$\pm$100 & 18.7\tablefootmark{d} & 4.545$\pm$1.446 & 40$\times$15\tablefootmark{c} & 15.0 & -2.44$\pm$0.21 & I, \du, rms: 6.8~mK & $n_e$: 12; $D$: 2; v: 25; CO: This work\\
NGC~2392 & 600 & 15.0 & 1.818$\pm$0.165 & 46$\times$44 & 120.0 & -2.34$\pm$0.09 & N, \du, rms: 57~mK & $n_e$,$T_e$: 26; $D$: 2; v: 6; CO: 4\\
NGC~5189 & 1550$\pm$200 & 11.60$\pm$0.28 & 1.471$\pm$0.043 & 163$\times$108 & 36.5 & -3.14$\pm$0.10 & S, \du, rms: 82~mK & $n_e$,$T_e$: 27; $D$: 2; v: 3; CO: 11\\
V458~Vul & 155 & 10\tablefootmark{a} & 12.5$\pm$2.0 & 27$\times$17 & 20.0 & -4.35$\pm$0.04 & I, \du, rms: 8.1~mK & $n_e$,v: 28; $D$: 19; CO: This work\\
\hline
\multicolumn{9}{c}{\textsc{Regular PNe}}\\
\hline
Abell~21 & 195$\pm$45 & 11.2 & 0.592$\pm$0.025 & 750$\times$515 & 45.0 & -4.70$\pm$0.06 & N, \du, rms: 90~mK & $n_e$: 29; $T_e$: 16; $D$: 2; v: 3; CO: 4\\
Abell~24 & 15$^{+60}_{-15}$ & 10.2 & 0.719$\pm$0.052 & 396$\times$360 & 14.0 & -5.04$\pm$0.06 & N, \du, rms: 60~mK & $n_e$: 30; $T_e$: 16; $D$: 2; v: 3; CO: 4\\
Abell~57 & 135 & 10\tablefootmark{a} & 2.128$\pm$0.317 & 40$\times$34 & 24.0 & -3.77$\pm$0.09 & A, \td, rms: 157~mK & $n_e$: 31; $D$: 2; v: 32; CO: 13\\
BD+303639 & 11000$\pm$1100 & 8.40$\pm$1.00 & 1.613$\pm$0.078 & 6.2$\times$5.6 & 35.5 & 0.12$\pm$0.08 & I, \du, $I$: 4.7~\kkms & $n_e$,$T_e$: 33; $D$: 2; v: 6; CO: 11\\
Cn~3-1 & 9480 & 7.7 & 7.143$\pm$1.531 & 5.7$\times$4.6 & 6.7 & -0.83$\pm$0.29 & I, \du, rms: 29~mK & $n_e$,$T_e$: 34; $D$: 2; v: 35; CO: 36\\
H~2-20 & 4800 & 7.0 & 8.3$\pm$2.4 & 2.8$\times$2.7 & 29.0 & -1.13$\pm$0.35 & I, \du, rms: 27~mK & $n_e$,$T_e$: 37; $D$: 19; v: 38; CO: 36\\
H~3-75 & 86$\pm$181 & 11.39$\pm$0.61 & 4.000$\pm$0.320 & 31$\times$30 & -- & -3.35$\pm$0.13 & N, \du, rms: 65~mK & $n_e$,$T_e$: 39; $D$: 2; CO: 4\\
Hen~1-2 & 4073 & 10.2 & 10.000$\pm$3.000 & 5$\times$5 & -- & -1.06$\pm$0.26 & I, \du, $I$: 5.0~\kkms & $n_e$: 20; $T_e$: 16; $D$: 2; CO: 36\\
Hen~2-36 & 600 & 16.5 & 4.000$\pm$0.160 & 24.8$\times$15.3 & 53.0 & -2.08$\pm$0.09 & S, \du, rms: 54~mK & $n_e$,v: 40; $T_e$: 16; $D$: 2; CO: 11\\
Hen~2-39 & 704$^{+95}_{-150}$ & 14.59$^{+1.00}_{-0.30}$ & 7.6$^{+1.5}_{-1.3}$ & 12.4$\times$12.2 & -- & -2.67$\pm$0.23 & A, \td, rms: 101~mK & $n_e$,$T_e$: 41; $D$: 19; CO: 13\\
Hen~2-68 & 14900$^{+49300}_{-6040}$ & 10.61$^{+0.72}_{-1.13}$ & 7.692$\pm$2.367 & 2.5$\times$2.5 & -- & -0.57$\pm$0.07 & A, \td, rms: 161~mK & $n_e$,$T_e$: 41; $D$: 2; CO: 13\\
Hen~2-113 & 4677$\pm$1077 & 10.2 & 2.083$\pm$0.130 & 1.5$\times$1.3 & 22.5 & 0.45$\pm$0.08 & C, \du, $F$: 8.2$\times$10$^{3}$~\kkmsa & $n_e$: 42; $T_e$: 16; $D$: 2; v: 43; CO: 11\\
Hen~2-131 & 7080$\pm$1630 & 20.40$^{+3.60}_{-3.19}$ & 2.703$\pm$0.219 & 10.0$\times$9.6 & 12.0 & -0.69$\pm$0.11 & A, \td, rms: 96~mK & $n_e$: 29; $T_e$: 41; $D$: 2; v: 3; CO: 13\\
Hen~2-459 & 16170$\pm$3230 & 10.00$\pm$1.00 & 1.010$\pm$0.306 & 3$\times$2 & -- & -0.34$\pm$0.49 & I, \du, $I$: 7.2~\kkms & $n_e$,$T_e$: 33; $D$: 2; CO: 36\\
Hen~3-1333 & 692$\pm$80 & 10.2 & 1.471$\pm$0.108 & 3.2$\times$2.8 & 31.6 & -1.16$\pm$0.28 & C, \du, $F$: 2.7$\times$10$^{4}$~\kkmsa & $n_e$: 42; $T_e$: 16; $D$: 2; v: 43; CO: 11\\
Hu~2-1 & 4073$\pm$938 & 9.9 & 2.381$\pm$0.397 & 8.0$\times$2.8 & 9.5 & -0.53$\pm$0.07 & A, \td, rms: 93~mK & $n_e$: 29; $T_e$: 34; $D$: 2; v: 3; CO: 13\\
IC~289 & 860 & 15.5 & 1.587$\pm$0.126 & 46$\times$44 & 25.5 & -2.82$\pm$0.20 & N, \du, rms: 270~mK & $n_e$: 15; $T_e$: 16; $D$: 2; v: 3; CO: 4\\
IC~418 & 12000$\pm$17000 & 9.1 & 1.370$\pm$0.056 & 14$\times$11 & 7.5 & -0.27$\pm$0.09 & I, \du, rms: 35~mK & $n_e$,$T_e$: 44; $D$: 2; v: 3; CO: 11\\
IC~972 & 1300 & 10.9 & 2.222$\pm$0.494 & 47$\times$47 & 16.0 & -4.09$\pm$0.09 & N, \du, rms: 54~mK & $n_e$: 45; $T_e$: 16; $D$: 2; v: 3; CO: 4\\
IC~1747 & 3930 & 10.9 & 3.846$\pm$0.592 & 13$\times$13 & 27.5 & -1.64$\pm$0.24 & I, \du, rms: 51~mK & $n_e$,$T_e$: 34; $D$: 2; v: 3; CO: 11\\
IC~2149 & 3827$\pm$2334 & 10.03$\pm$0.48 & 1.852$\pm$0.137 & 12.5$\times$8.0 & 24.0 & -1.08$\pm$0.07 & I, \du, rms: 71~mK & $n_e$,$T_e$: 39; $D$: 2; v: 6; CO: 11\\
IC~3568 & 1900 & 11.4 & 2.273$\pm$0.207 & 17.8$\times$17.8 & 8.0 & -1.94$\pm$0.06 & N, \du, rms: 63~mK & $n_e$,$T_e$: 46; $D$: 2; v: 3; CO: 4\\
IC~4406 & 1350$\pm$310 & 10.50$\pm$1.05 & 1.136$\pm$0.155 & 46.4$\times$29.9 & 7.0 & -2.47$\pm$0.07 & S, \du, $F$: 5.1$\times$10$^{4}$~\kkmsa & $n_e$,$T_e$: 29; $D$: 2; v: 3; CO: 11\\
IC~4593 & 3236$^{+1640}_{-1490}$ & 12.6 & 2.632$\pm$0.346 & 15.3$\times$14.7 & 15.0 & -1.64$\pm$0.06 & N, \du, rms: 81~mK & $n_e$: 42; $T_e$: 47; $D$: 2; v: 6; CO: 4\\
IC~4732 & 12500 & 13.0 & 8.3$\pm$2.4 & 1.4$\times$1.4 & 20.0 & -0.15$\pm$0.11 & I, \du, rms: 24~mK & $n_e$,$T_e$: 48; $D$: 19; v: 49; CO: 36\\
IRAS~21282 & 1900 & 10\tablefootmark{a} & 3.704$\pm$0.274 & 6.0$\times$4.5 & 14.8 & -0.8$\pm$0.34 & I, \du, $I$: 279.0~\kkms & $n_e$: 50; $D$: 2; v: 17; CO: 11\\
J~900 & 3980$^{+790}_{-650}$ & 12.00$^{+0.04}_{-0.02}$ & 4.55$\pm$0.25 & 8.2$\times$7.8 & 18.0 & -1.3$\pm$0.13 & I, \du, $I$: 0.9~\kkms & $n_e$,$T_e$: 1; $D$: 19; v: 3; CO: 11\\
Jn~1 & 14 & 12.5 & 0.990$\pm$0.069 & 354$\times$298 & 15.0 & -4.95$\pm$0.09 & N, \du, rms: 130~mK & $n_e$: 29; $T_e$: 16; $D$: 2; v: 3; CO: 4\\
JnEr~1 & 10 & 10.60$^{+0.90}_{-0.60}$ & 0.943$\pm$0.071 & 394$\times$345 & 22.5 & -5.06$\pm$0.09 & I, \du, $F$: 1.0$\times$10$^{4}$~\kkmsa & $n_e$,$T_e$: 51; $D$: 2; v: 3; CO: 11\\
K~3-35 & 24660 & 17.0 & 3.9$^{+0.7}_{-0.5}$ & 6$\times$3 & 10.0 & -1.74$\pm$0.37 & I, \du, $I$: 5.7~\kkms & $n_e$: 52; $T_e$: 53; $D$: 19; v: 6; CO: 36\\
K~3-72 & 185$^{+75}_{-65}$ & 10.20$^{+0.70}_{-0.50}$ & 4.6$\pm$0.8 & 22.9$\times$18.0 & 13.0 & -3.48$\pm$0.22 & I, \du, rms: 48~mK & $n_e$,$T_e$: 51; $D$: 19; v: 3; CO: 36\\
M~1-4 & 3596$\pm$2080 & 11.84$\pm$0.67 & 3.3$\pm$0.35 & 4.2$\times$4.2 & 13.5 & -0.68$\pm$0.16 & I, \du, rms: 100~mK & $n_e$,$T_e$: 39; $D$: 19; v: 3; CO: 11\\
M~1-6 & 19530$\pm$39320 & 7.90$\pm$0.43 & 7.692$\pm$2.367 & 4.0$\times$2.7 & 24.0 & -0.3$\pm$0.31 & I, \du, rms: 65~mK & $n_e$,$T_e$: 39; $D$: 2; v: 54; CO: 11\\
M~1-11 & 12000$^{+6600}_{-3500}$ & 10.00$\pm$1.58 & 4.545$\pm$0.620 & 5.2$\times$5.1 & -- & -0.52$\pm$0.19 & A, \td, rms: 113~mK & $n_e$: 1; $T_e$: 39; $D$: 2; CO: 13\\
M~1-12 & 7943$\pm$1829 & 10.2 & 4.545$\pm$0.620 & 1.8$\times$1.8 & -- & -0.11$\pm$0.23 & I, \du, rms: 150~mK & $n_e$,$T_e$: 29; $D$: 2; CO: 11\\
M~1-32 & 8350 & 9.43$\pm$0.22 & 2.632$\pm$0.416 & 9.1$\times$8.0 & 15.0 & -1.21$\pm$0.18 & I, \du, $I$: 0.8~\kkms & $n_e$,$T_e$: 55; $D$: 2; v: 6; CO: 36\\
M~1-46 & 3700$\pm$90 & 10.2 & 2.381$\pm$0.113 & 12.1$\times$11.3 & 7.0 & -1.27$\pm$0.38 & I, \du, rms: 42~mK & $n_e$,$T_e$: 29; $D$: 2; v: 3; CO: 36\\
M~1-65 & 4170 & 10.2 & 6.667$\pm$0.889 & 4.2$\times$4.0 & 4.0 & -1.08$\pm$0.13 & I, \du, rms: 60~mK & $n_e$: 56; $T_e$: 16; $D$: 2; v: 3; CO: 36\\
M~1-71 & 4570 & 9.80$\pm$0.50 & 2.9$\pm$0.4 & 6.0$\times$3.7 & 16.5 & 0.06$\pm$0.21 & I, \du, rms: 51~mK & $n_e$,$T_e$: 20; $D$: 19; v: 3; CO: 36\\
M~1-73 & 6130 & 7.4 & 4.545$\pm$0.620 & 8.8$\times$6.0 & 11.0 & -1.21$\pm$0.17 & I, \du, rms: 24~mK & $n_e$,$T_e$: 34; $D$: 2; v: 3; CO: 36\\
M~2-53 & 480$^{+180}_{-90}$ & 11.70$^{+0.50}_{-0.40}$ & 6.0$\pm$1.0 & 20$\times$15 & 11.0 & -2.87$\pm$0.15 & I, \du, $I$: 7.8~\kkms & $n_e$,$T_e$: 51; $D$: 19; v: 3; CO: 36\\
M~2-55 & 510$\pm$130 & 10.2 & 0.658$\pm$0.022 & 58$\times$40 & -- & -3.36$\pm$0.1 & I, \du, rms: 71~mK & $n_e$: 29; $T_e$: 16; $D$: 2; CO: 11\\
M~3-3 & 330$\pm$40 & 12.60$^{+0.08}_{-0.05}$ & 5.5$^{+1.3}_{-1.8}$ & 16.6$\times$15.8 & 10.0 & -3.23$\pm$0.09 & I, \du, $I$: 6.6~\kkms & $n_e$,$T_e$: 1; $D$: 19; v: 3; CO: 11\\
M~3-28 & 100 & 10.8 & 2.5$^{+1.1}_{-1.3}$ & 24.1$\times$12.1 & -- & -2.32$\pm$0.21 & I, \du, $I$: 147.1~\kkms & $n_e$,$T_e$: 57; $D$: 19; CO: 36\\
M~3-35 & 7244$\pm$1668 & 6.4 & 1.000$\pm$0.310 & 4.6$\times$4.0 & 24.5 & -0.2$\pm$0.24 & N, \du, rms: 110~mK & $n_e$: 29; $T_e$: 58; $D$: 2; v: 3; CO: 4\\
M~4-18 & 11360$\pm$13600 & 10\tablefootmark{a} & 6.667$\pm$0.889 & 3.7$\times$3.5 & 17.0 & -1.06$\pm$0.13 & I, \du, rms: 54~mK & $n_e$,$T_e$: 39; $D$: 2; v: 3; CO: 11\\
Me~1-1 & 6760$^{+1560}_{-1400}$ & 10.2 & 3.704$\pm$0.274 & 6.0$\times$2.8 & 9.0 & -0.92$\pm$0.17 & I, \du, rms: 25~mK & $n_e$: 42; $T_e$: 29; $D$: 2; v: 3; CO: 36\\
Mz~1 & 400 & 11.3 & 1.266$\pm$0.208 & 49.3$\times$35.3 & 6.0 & -2.72$\pm$0.14 & S, \du, $I$: 4.8~\kkms & $n_e$: 45; $T_e$: 16; $D$: 2; v: 3; CO: 11\\
NGC~40 & 1738 & 10.6 & 1.786$\pm$0.064 & 56$\times$34 & 29.0 & -2.25$\pm$0.08 & I, \du, rms: 87~mK & $n_e$,$T_e$: 46; $D$: 2; v: 3; CO: 11\\
NGC~1501 & 823$\pm$423 & 10.83$\pm$0.56 & 1.724$\pm$0.059 & 57$\times$50 & 37.0 & -2.42$\pm$0.17 & N, \du, rms: 150~mK & $n_e$,$T_e$: 39; $D$: 2; v: 3; CO: 4\\
NGC~1535 & 2500 & 11.2 & 1.370$\pm$0.113 & 33.3$\times$32.1 & 20.0 & -2.23$\pm$0.06 & N, \du, rms: 69~mK & $n_e$,$T_e$: 59; $D$: 2; v: 3; CO: 4\\
NGC~2022 & 1500 & 15.0 & 2.273$\pm$0.258 & 27.9$\times$25.5 & 26.0 & -2.51$\pm$0.07 & N, \du, rms: 140~mK & $n_e$,$T_e$: 60; $D$: 2; v: 3; CO: 4\\
NGC~2371 & 230 & 13.40$\pm$1.00 & 1.724$\pm$0.149 & 48.9$\times$30.6 & 42.5 & -2.91$\pm$0.11 & I, \du, rms: 74~mK & $n_e$,$T_e$: 1; $D$: 2; v: 3; CO: 11\\
NGC~2438 & 250 & 10.7 & 0.725$\pm$0.116 & 80.7$\times$78.3 & 22.5 & -3.40$\pm$0.08 & I, \du, rms: 88~mK & $n_e$,$T_e$: 61; $D$: 2; v: 3; CO: 11\\
NGC~2440 & 6000 & 16.1 & 1.77$\pm$0.45 & 58.9$\times$25.1 & 22.5 & -1.99$\pm$0.10 & I, \du, $F$: 4.7$\times$10$^{3}$~\kkmsa & $n_e$,$T_e$: 60; $D$: 19; v: 3; CO: 11\\
NGC~2452 & 1820$\pm$80 & 12.6 & 2.941$\pm$0.692 & 18.3$\times$12.4 & 32.0 & -1.99$\pm$0.07 & I, \du, rms: 55~mK & $n_e$,$T_e$: 1; $D$: 2; v: 3; CO: 11\\
NGC~2818 & 1000$\pm$140 & 15.0 & 3.0$\pm$0.8 & 56.2$\times$46.0 & 36.5 & -3.24$\pm$0.10 & S, \du, $F$: 1.6$\times$10$^{3}$~\kkmsa & $n_e$,$T_e$: 44; $D$: 19; v: 3; CO: 11\\
NGC~2899 & 130$\pm$20 & 19.50$\pm$2.00 & 1.923$\pm$0.111 & 68.5$\times$59.8 & 25.0 & -2.96$\pm$0.08 & S, \du, $F$: 7.0$\times$10$^{3}$~\kkmsa & $n_e$,$T_e$: 1; $D$: 2; v: 3; CO: 11\\
NGC~3132 & 600 & 9.5 & 0.758$\pm$0.017 & 86$\times$60 & 14.0 & -2.75$\pm$0.06 & S, \du, $F$: 4.0$\times$10$^{4}$~\kkmsa & $n_e$,$T_e$: 60; $D$: 2; v: 3; CO: 11\\
NGC~3242 & 2000 & 11.7 & 1.333$\pm$0.089 & 45$\times$39 & 27.5 & -1.76$\pm$0.06 & N, \du, rms: 59~mK & $n_e$,$T_e$: 60; $D$: 2; v: 6; CO: 4\\
NGC~3587 & 214$\pm$49 & 10.90$\pm$1.09 & 0.813$\pm$0.033 & 208$\times$202 & 26.0 & -3.85$\pm$0.06 & N, \du, rms: 95~mK & $n_e$,$T_e$: 29; $D$: 2; v: 3; CO: 4\\
NGC~3699 & 560 & 19.0 & 1.370$\pm$0.150 & 47$\times$37 & 27.5 & -2.94$\pm$0.12 & S, \du, rms: 58~mK & $n_e$,$T_e$: 47; $D$: 2; v: 3; CO: 11\\
NGC~3918 & 4370$^{+1700}_{-850}$ & 12.73$\pm$0.40 & 4.545$\pm$1.446 & 18.7$\times$17.1 & 22.6 & -1.07$\pm$0.09 & A, \td, rms: 92~mK & $n_e$,$T_e$: 41; $D$: 2; v: 6; CO: 13\\
NGC~4361 & 1200$\pm$200 & 17.90$\pm$0.30 & 1.031$\pm$0.043 & 119$\times$115 & 32.0 & -3.47$\pm$0.06 & N, \du, rms: 30~mK & $n_e$,$T_e$: 62; $D$: 2; v: 3; CO: 4\\
NGC~5315 & 10000 & 9.0 & 0.962$\pm$0.185 & 10.7$\times$9.2 & 36.0 & -0.56$\pm$0.12 & S, \du, rms: 46~mK & $n_e$,$T_e$: 60; $D$: 2; v: 3; CO: 11\\
NGC~6058 & 1410$^{+810}_{-680}$ & 13.2 & 2.778$\pm$0.231 & 36$\times$28 & 27.5 & -3.58$\pm$0.04 & I, \du, rms: 58~mK & $n_e$: 42; $T_e$: 16; $D$: 2; v: 3; CO: 11\\
NGC~6072 & 390$\pm$54 & 11.4 & 0.917$\pm$0.168 & 74.3$\times$65.1 & 17.0 & -2.81$\pm$0.09 & S, \du, $F$: 9.3$\times$10$^{4}$~\kkmsa & $n_e$: 42; $T_e$: 16; $D$: 2; v: 17; CO: 11\\
NGC~6210 & 4365 & 9.7 & 2.041$\pm$0.125 & 14$\times$14 & 34.2 & -1.12$\pm$0.08 & I, \du, rms: 42~mK & $n_e$,$T_e$: 46; $D$: 2; v: 6; CO: 11\\
NGC~6302 & 14000 & 18.4 & 1.17$\pm$0.14 & 90$\times$35 & 12.0 & -1.48$\pm$0.10 & N, \du, $I$: 19.9~\kkms & $n_e$,$T_e$: 60; $D$: 19; v: 63; CO: 64\\
NGC~6309 & 2400 & 11.3 & 2.632$\pm$0.416 & 22.8$\times$12.4 & 34.0 & -1.83$\pm$0.12 & I, \du, rms: 66~mK & $n_e$: 20; $T_e$: 16; $D$: 2; v: 3; CO: 11\\
NGC~6369 & 3550$\pm$1130 & 10.65$\pm$0.23 & 1.087$\pm$0.059 & 30$\times$29 & 41.5 & -1.01$\pm$0.17 & I, \du, rms: 172~mK & $n_e$,$T_e$: 55; $D$: 2; v: 3; CO: 11\\
NGC~6543 & 6460 & 7.9 & 1.370$\pm$0.056 & 26.5$\times$23.5 & 16.0 & -1.12$\pm$0.05 & I, \du, rms: 210~mK & $n_e$,$T_e$: 55; $D$: 2; v: 6; CO: 11\\
NGC~6563 & 134$\pm$70 & 10.74$\pm$0.47 & 0.935$\pm$0.114 & 59$\times$43 & 21.5 & -3.05$\pm$0.07 & S, \du, $F$: 4.8$\times$10$^{4}$~\kkmsa & $n_e$,$T_e$: 65; $D$: 2; v: 17; CO: 11\\
NGC~6572 & 25700 & 10.6 & 1.852$\pm$0.206 & 15$\times$13 & 14.0 & -0.58$\pm$0.09 & I, \du, rms: 161~mK & $n_e$,$T_e$: 46; $D$: 2; v: 6; CO: 11\\
NGC~6629 & 1380 & 8.8 & 2.041$\pm$0.083 & 16.6$\times$15.5 & 12.0 & -1.29$\pm$0.11 & I, \du, rms: 85~mK & $n_e$: 20; $T_e$: 16; $D$: 2; v: 3; CO: 11\\
NGC~6644 & 15000$\pm$2000 & 12.80$\pm$0.03 & 8.3$\pm$2.4 & 4.4$\times$4.3 & -- & -0.66$\pm$0.13 & A, \td, rms: 114~mK & $n_e$,$T_e$: 66; $D$: 19; CO: 13\\
NGC~6720 & 500 & 10.6 & 0.787$\pm$0.025 & 89$\times$66 & 26.5 & -2.54$\pm$0.09 & I, \du, $F$: 7.6$\times$10$^{4}$~\kkmsa & $n_e$,$T_e$: 46; $D$: 2; v: 3; CO: 11\\
NGC~6741 & 2000 & 12.6 & 2.6$\pm$0.55 & 9.1$\times$6.5 & 23.4 & -0.92$\pm$0.20 & I, \du, rms: 116~mK & $n_e$,$T_e$: 46; $D$: 19; v: 6; CO: 11\\
NGC~6772 & 230 & 11.7 & 0.901$\pm$0.146 & 80.7$\times$70.8 & 11.0 & -3.07$\pm$0.12 & I, \du, $F$: 2.0$\times$10$^{4}$~\kkmsa & $n_e$: 67; $T_e$: 16; $D$: 2; v: 3; CO: 11\\
NGC~6781 & 500 & 9.0 & 0.500$\pm$0.018 & 180$\times$109 & 12.0 & -2.99$\pm$0.1 & I, \du, $F$: 4.0$\times$10$^{5}$~\kkmsa & $n_e$,$T_e$: 68; $D$: 2; v: 6; CO: 11\\
NGC~6826 & 5130 & 9.4 & 1.299$\pm$0.067 & 27$\times$24 & 6.0 & -1.46$\pm$0.08 & S, \du, rms: 87~mK & $n_e$,$T_e$: 46; $D$: 2; v: 3; CO: 4\\
NGC~6853 & 95$\pm$80 & 10.90$\pm$0.57 & 0.389$\pm$0.006 & 475$\times$340 & 30.0 & -3.43$\pm$0.07 & I, \du, $F$: 4.6$\times$10$^{4}$~\kkmsa & $n_e$,$T_e$: 65; $D$: 2; v: 3; CO: 11\\
NGC~6884 & 8130$^{+2060}_{-1870}$ & 9.4 & 3.3$\pm$1.24 & 7.5$\times$7.0 & 19.0 & -0.79$\pm$0.08 & I, \du, rms: 20~mK & $n_e$: 42; $T_e$: 44; $D$: 19; v: 6; CO: 36\\
NGC~6894 & 29900 & 8.2 & 1.449$\pm$0.231 & 56.4$\times$53.3 & 43.0 & -2.77$\pm$0.08 & N, \du, rms: 170~mK & $n_e$: 69; $T_e$: 58; $D$: 2; v: 3; CO: 4\\
NGC~7008 & 1175$\pm$160 & 12.2 & 0.645$\pm$0.033 & 99.0$\times$81.5 & 40.0 & -2.94$\pm$0.1 & I, \du, $F$: 1.6$\times$10$^{3}$~\kkmsa & $n_e$: 42; $T_e$: 20; $D$: 2; v: 3; CO: 11\\
NGC~7009 & 3164$\pm$1683 & 10.12$\pm$0.76 & 1.235$\pm$0.091 & 28$\times$22 & 20.8 & -1.25$\pm$0.07 & N, \du, rms: 83~mK & $n_e$,$T_e$: 65; $D$: 2; v: 6; CO: 4\\
NGC~7026 & 5510 & 9.3 & 3.226$\pm$0.312 & 39$\times$18 & 38.0 & -1.80$\pm$0.08 & N, \du, rms: 55~mK & $n_e$,$T_e$: 34; $D$: 2; v: 3; CO: 4\\
NGC~7027 & 47000 & 14.0 & 0.92$\pm$0.1 & 15.6$\times$12.0 & 21.5 & 0.14$\pm$0.09 & N, \du, $F$: 1.0$\times$10$^{0}$~\kkmsa & $n_e$,$T_e$: 70; $D$: 19; v: 3; CO: 71\\
NGC~7048 & 1780 & 12.6 & 1.587$\pm$0.529 & 63$\times$60 & 15.0 & -3.26$\pm$0.13 & I, \du, rms: 149~mK & $n_e$: 72; $T_e$: 16; $D$: 2; v: 3; CO: 11\\
NGC~7354 & 7950 & 12.2 & 2.083$\pm$0.304 & 33$\times$31 & 28.0 & -1.65$\pm$0.13 & N, \du, $F$: 1.0$\times$10$^{0}$~\kkmsa & $n_e$: 29; $T_e$: 16; $D$: 2; v: 6; CO: 73\\
NGC~7662 & 3300 & 13.4 & 1.754$\pm$0.092 & 30.5$\times$28.0 & 25.0 & -1.63$\pm$0.06 & I, \du, rms: 52~mK & $n_e$,$T_e$: 46; $D$: 2; v: 6; CO: 11\\
PB~1 & 2000$\pm$7700 & 12.17$\pm$0.69 & 3.226$\pm$0.937 & 10.6$\times$9.5 & 20.0 & -2.28$\pm$0.11 & A, \td, rms: 112~mK & $n_e$: 1; $T_e$: 39; $D$: 2; v: 74; CO: 13\\
Sh~2-71 & 324$\pm$75 & 14.20$\pm$1.42 & 1.613$\pm$0.052 & 132.4$\times$74.9 & 14.0 & -3.51$\pm$0.31 & N, \du, rms: 200~mK & $n_e$,$T_e$: 29; $D$: 2; v: 3; CO: 4\\
SwSt~1 & 16200 & 7.5 & 2.941$\pm$0.952 & 5.6$\times$5.2 & 9.0 & -0.42$\pm$0.07 & I, \du, rms: 102~mK & $n_e$,$T_e$: 69; $D$: 2; v: 3; CO: 11\\
Vy~1-1 & 3390 & 10.2 & 5.000$\pm$0.750 & 5.2$\times$5.2 & 10.0 & -1.46$\pm$0.12 & N, \du, rms: 74~mK & $n_e$: 20; $T_e$: 16; $D$: 2; v: 3; CO: 4\\
Vy~2-2 & 11730 & 13.9 & 3.5$\pm$1.2 & 3.1$\times$2.6 & 17.5 & 0.3$\pm$0.21 & C, \du, $F$: 8.1$\times$10$^{2}$~\kkmsa & $n_e$,$T_e$: 34; $D$: 19; v: 3; CO: 11\\
Vy~2-3 & 2800$\pm$650 & 10.3 & 6.250$\pm$1.172 & 4.6$\times$4.6 & 14.0 & -1.53$\pm$0.08 & I, \du, $I$: 0.4~\kkms & $n_e$,$T_e$: 29; $D$: 2; v: 3; CO: 36\\
\hline
\end{longtable}
\tablefoot{
Telescope code: N:NRAO 12m; I: IRAM~30m; S: SEST~15m; A: APEX~12m; C: CSO~10m.
\tablefoottext{a}{Assumed value. }
\tablefoottext{b}{Size and H$\alpha$ flux from \citet{corradi11b}.}
\tablefoottext{c}{Size used for ionised mass determination. Estimate of molecular mass used a smaller size based on the equatorial ring, 14$\times$7.5 arcsec$^2$ (see section~\ref{ngc6778}).}
\tablefoottext{d}{Based on data presented in \citet{wesson18}.}
}
\tablebib{
Every size (diameters) and $S_0$(H$\alpha$) values from \citet{frew16}, unless specified otherwise. (1) \citet{kingsburgh94}; (2) \citet{gaia20}; (3) \citet{weinberger89}; (4) \citet{huggins89}; (5) \citet{corradi15}; (6) \citet{guerrero20}; (7) \citet{corradi11}; (8) \citet{miszalski11b}; (9) \citet{munday20}; (10) \citet{jones15}; (11) \citet{huggins96}; (12) \citet{wesson18}; (13) \citet{guzman18}; (14) \citet{tsamis03}; (15) \citet{kaler70}; (16) \citet{cahn92}; (17) \citet{gussie94}; (18) \citet{jones16}; (19) \citet{frew16}; (20) \citet{stanghellini89}; (21) \citet{odell98}; (22) \citet{corradi14}; (23) \citet{miszalski19}; (24) \citet{jones10b}; (25) \citet{rodriguez01}; (26) \citet{phillips99}; (27) \citet{garciarojas12}; (28) \citet{wesson08}; (29) \citet{phillips98}; (30) \citet{bohigas03}; (31) \citet{abell66}; (32) \citet{pereyra13}; (33) \citet{sterling08}; (34) \citet{wesson05}; (35) \citet{gussie89}; (36) \citet{huggins05}; (37) \citet{pottasch15}; (38) \citet{gesicki14}; (39) \citet{henry10}; (40) \citet{corradi93}; (41) \citet{gorny14}; (42) \citet{wang04}; (43) \citet{danehkar15}; (44) \citet{torres-peimbert77}; (45) \citet{perinotto94}; (46) \citet{liu04}; (47) \citet{costa96}; (48) \citet{aller87}; (49) \citet{gesicki00}; (50) \citet{hsia19}; (51) \citet{bohigas01}; (52) \citet{miranda00}; (53) \citet{aaquist93}; (54) \citet{gesicki06}; (55) \citet{garcia-rojas12}; (56) \citet{kondrateva79}; (57) \citet{kaler96}; (58) \citet{mckenna96}; (59) \citet{pottasch11}; (60) \citet{tsamis04}; (61) \citet{ottl14}; (62) \citet{liu98}; (63) \citet{szyszka11}; (64) \citet{santander17}; (65) \citet{krabbe05}; (66) \citet{barker78}; (67) \citet{meatheringham98}; (68) \citet{mavromatakis01}; (69) \citet{milanova09}; (70) \citet{zhang05}; (71) \citet{santander12}; (72) \citet{gurzadyan97}; (73) Verbena et al. (in preparation); (74) \citet{bandyopadhyay20}
}
\end{landscape}

\end{appendix}

\end{document}